\documentclass[fleqn,usenatbib]{mnras}

\usepackage{newtxtext,newtxmath}

\usepackage[T1]{fontenc}

\DeclareRobustCommand{\VAN}[3]{#2}
\let\VANthebibliography\thebibliography
\def\thebibliography{\DeclareRobustCommand{\VAN}[3]{##3}\VANthebibliography}

\usepackage{color}
\usepackage{soul}
\usepackage{graphicx}	
\usepackage{amsmath}	

\newcommand{\comment}[1]{}

\newcommand{\2}{$_{2}$}

\newcommand{\4}{$_{4}$}

\newcommand{\tnu}{\tilde{\nu}}
\newcommand{\tE}{\tilde{E}}

\newcommand{\cm}{cm$^{-1}$}
\newcommand{\um}{$\mu$m}

\title[Non-LTE spectra of molecules for exoplanets]{Non-Local thermal equilibrium spectra of atmospheric molecules for exoplanets}

\author[S. O. M. Wright et al.]{
Sam O. M. Wright,$^{1}$\thanks{E-mail: swright@star.ucl.ac.uk}
Ingo Waldmann,$^{1}$
and Sergei N. Yurchenko$^{1}$
\\
$^{1}$Department of Physics and Astronomy, University College London, Gower Street, WC1E 6BT London, United Kingdom
}

\date{Accepted 2022 March 2. Received 2022 February 2; in original form 2021 December 12}

\pubyear{2021}

\begin{document}
\label{firstpage}
\pagerange{\pageref{firstpage}--\pageref{lastpage}}
\maketitle

\begin{abstract}
Here we present a study of non-LTE effects on the exoplanetary spectra of a collection of molecules which are key in the investigation of exoplanet atmospheres: water, methane, carbon monoxide and titanium oxide. These molecules are chosen as examples of different spectral ranges (IR and UV), molecular types (diatomics and polyatomics) and spectral types (electronic and ro-vibrational); the importance of different vibrational bands in forming distinct non-LTE spectral features are investigated. 
Most notably, such  key spectral signatures for distinguishing between the LTE and non-LTE cases include: for CH\4\ the 3.15 ~\um\ band region; for H\2O  the 2.0~\um\ and 2.7~\um\  band regions; for TiO,  a strong variation in intensity in the bands between 0.5 and 0.75 \um; and a sole CO signature between 5 and 6 \um. The analysis is based on the ExoMol cross sections and takes advantage of the extensive vibrational assignment of these molecular line lists in the ExoMol database. We examine LTE and non-LTE cross sections under conditions consistent with those on WASP-12b and WASP-76b using the empirically motivated bi-temperature Treanor model. In addition, we make a simplistic forward model simulation of transmission spectra for H\2O  in the atmosphere of WASP-12b  using the TauREx 3 atmospheric modelling code.

\end{abstract}




\section{Introduction}

\label{sec:intro}
While much of the study into exoplanet atmospheres assumes species to be in local thermodynamic equilibrium, effects arising from non-local thermodynamic equilibrium (non-LTE or NLTE) are known to be present and have been detected in Earth's atmosphere, as well as the atmospheres of other solar system planets \citep{01LoTaxx}, including Venus \citep{07LoDrCa}, Mars \citep{05LoLoLo} and the gas giants \citep{15KiSiHo}, in stellar atmospheres, comets \citep{84WeMuxx} and the ISM  \citep{99GoLaxx.ISM,07vaBlSc,09LivaKl.NO}.
Non-LTE spectroscopy in general can be traced to \citet{30Milnex} followed by many key papers \citep{56CuGoxx, 69Hought, 69KuLoxx, 72Dickin, 74KuJaxx, 75Shvedx, 86LoRoLo, 86LoRoMo, 92WiPiSh}.

In addition to the importance of non-LTE effects in the spectroscopy of planetary atmospheres \citep{01LoTaxx}, they have been significant in the study of stellar atmospheres for quite some time,  including molecular detections such as that of Aluminium Oxide in the optical spectrum of VY Canis Majoris \citep{13KaScMe.AlO}, as well as TiO and TiO\2 \citep{13KaGoMe.TiO}. Non-LTE considerations have also played a role in constraining the abundance of SiO, CO and HCN in the stellar envelope of R Doradus \citep{18VaDeLo.nLTE}. The roles of non-LTE effects in atomic spectra have been investigated for exoplanet atmospheres \citep{19FiHexx} as well as for the ion H$_3^+$ in giant planets within our solar system \citep{19Drxxxx.nLTE} and extrasolar hot Jupiters \citep{07KoAyMi.nLTE}. They have also been hypothesized to be a factor in the molecular spectroscopy of exoplanet atmospheres \citep{11WaTiDr,10SwDeGr.exo}, although this has been disputed subsequently \citep{11MaDeBl}. Despite this, the non-LTE spectroscopy for complex molecules in exoplanet atmospheres has not been characterised; an initial characterisation of these effects in a number of the most commonly observed molecules is reported here. 

A molecule is considered to be in local thermodynamic equilibrium where it has equal energy in each of its modes of motion; those of vibration, translation and rotation. In local thermodynamic equilibrium a single temperature describes (i) the velocities of molecules (via the Maxwellian distribution), (ii) state populations (via the Boltzmann distribution) and (iii) ionization (via the Saha equation). In non-local thermodynamic equilibrium however, this does not hold and the molecules have differing energies across their different modes. In this non-LTE regime the assumption that energy level populations can be described by the Boltzmann distribution is no longer valid. This non-LTE condition prevails where the rate of collision between molecules is insufficient to drive the molecules back to a state of local thermodynamic equilibrium (i.e. with comparable energies across modes); at these collision rates, the time between collisions is greater than the time taken for de-excitation of the molecule from its non-equilibrated states. Such low collision rate conditions can be found in low pressure regions, typically high in a planet's atmosphere \citep{01LoTaxx}. For instance, CH$_{4}$ has been shown to exist in increasingly greater non-LTE for decreasing pressures below 0.1 mbar in gas giant atmospheres \citep{90Apxxxx}.

In this paper we examine the effects of spectral signatures indicating non-LTE populations in the atmospheres of exoplanets. To do this, we make use of the ExoCross software's \citep{ExoCross} capability to approximate cross sections for molecules not in local thermodynamic equilibrium in conjunction with the ExoMol molecular  database \citep{jt810}. 

When treated rigorously, non-LTE radiative transfer calculations solve  the locally defined statistical equilibrium equations and compute the non-LTE populations of all states considered \citep{02StvoFu.nLTE, GRANADA}.  These calculations require a set of spectroscopic and collisional data for all molecules and states involved, which are usually only partially available. This significantly limits their  application in terms of temperatures and frequencies observed.   It should be noted that typical modern exoplanetary atmospheric applications cannot cope with such limitations due to large temperature ranges and  comparatively low spectral resolution for space-borne observations, unsuited to populating the many parameters which are a requirement of full non-LTE population modelling. Approaching this with high-resolution spectroscopy introduces an additional limitation in discerning absolute line strengths - an important factor in distinguishing the LTE and non-LTE cases for molecular spectra. This work is an attempt to introduce a simplified, empirically motivated description of non-LTE effects in exoplanetary atmospheric retrievals. 
To this end, we present a detailed investigation of the contributions of different vibrational bands on forming distinct non-LTE spectral signatures for a few key molecules. As an example of such a parametrization, we explore a  bi-temperature model (vibrational and rotational)  sometimes  referred to as Treanor distribution~\citep{68TrRiRe.nLTE} to introduce non-LTE effects into radiative transfer calculations of exoplanets, at the molecular cross section level. 
The vibrational and rotational degrees of freedom are assumed to be in their LTE (Boltzmann equilibrium) \citep{19PaLaxx}, which is a common treatment  in many non-LTE  spectroscopic applications. Here it is common to assume the rotational degrees of freedom to be in LTE, with the non-LTE effects originating from non-Boltzmann  vibrational  populations obtained from statistical equilibrium equations. Such empirical evaluations are conducted in the lab, for instance using a shock region arising from a shaped nozzle, as in \citep{20DuSuBr.CO}. This approximation eliminates the dependence on specific non-LTE state population ratios that could hinder achieving the aim of a comprehensive non-LTE retrieval of an exoplanet atmosphere in further work. Such a Boltzmann-like parametrization can be considered as a natural first order approximation to the inclusion of the non-LTE populations, which  would otherwise require a large amount of a priori information as input to a non-LTE radiative transfer solver, see, e.g. \citep{GRANADA}. Naturally the ground truths of such information are not constrained for exoplanets and the impact of seasonal and diurnal variations adds additional granularity beyond current and near-future capabilities. Instead, we approach the problem with the two variable, bi-temperature parameterisation with the non-LTE state populations having a Boltzmann-like dependence. In this way, we average most non-LTE effects into the molecular opacity tables which can then be included in retrievals.

The non-LTE effects are considered for the example of four molecules of major atmospheric importance, H\2O, CH\4, CO and TiO,  where we take advantage of the extensive vibrational assignment provided in their respective  ExoMol line lists. Our formulation of non-LTE populations as molecular cross sections approximated in this way allows us to make non-LTE spectral absorption and emission effects available to existing retrieval frameworks with little to no modification.

\section{Method}\label{sec:method}

To begin characterising the effects of molecules in non-LTE within exoplanet atmospheres on spectra, we generated absorption cross sections for molecules using some assumed populations $F_{J,\varv,k}$  of the lower states $J,\varv,k$. For the LTE case, these are in accordance with the Boltzmann distributions for a given temperature $T$, as given by
\begin{equation}
\label{e:F:LTE}
F_{J,\varv,k}(T) = \frac{ g_{J,k,\varv}^{\rm tot} \,   e^{-c_2 \tE_{J,\varv,k}/T} }{Q(T)},
\end{equation}
where  $c_2= hc / k_B$ is the second radiation constant (cm K), $\tilde{E}_i = E_i/h c $ is the energy term value (\cm), $T$ is the temperature in K and  $g_{J,k,\varv}^{\rm tot}$ is  the total degeneracy 
$$
 g_{J,k,\varv}^{\rm tot} = g^{\rm ns}_{\varv,k} (2 J+1),
$$
 $Q(T)$ is the LTE partition function defined as a sum over states
\begin{equation}
\label{e:pf}
  Q(T) =\sum_{n}  g_{J,k,\varv}^{\rm tot} e^{-c_2 \tE_{J,\varv,k}/T};
\end{equation}
$J$ is  the corresponding total angular momentum and $g^{\rm ns}_{J,\varv,k}$ is the nuclear-spin statistical weight factor,  $\varv$ and $k$ are generic vibrational (vibronic) and rotational quantum numbers, respectively.

For the non-LTE case we will assume that the rotational states can also be populated by the Boltzmann distribution corresponding to some rotational temperature $T_{\rm rot}$, which is also the kinetic temperature of the environment \citep{01LoTaxx} and the total population of $J,\varv,k$ is then approximated by a product 
\begin{equation}
\label{e:F:non-LTE:1}
F_{J,\varv,k}(T_{\rm rot}) = \frac{ g_{J,k,\varv}^{\rm tot} \, F_{\varv} e^{-c_2 \tE_{J,k}^{\varv,\rm rot}/T_{\rm rot}}}{Q_{\rm rot}(T)}
\end{equation}
with some non-LTE vibrational populations $F_{\varv}$. These non-LTE cross sections are created using the ExoCross software \citep{ExoCross} with ExoMol line lists \citep{jt810}. Here the rotational energies are estimated using the following factorization 
\begin{equation}\label{e:Erot:Evib}
  \tE_{\varv,J,k} = \tE_{\varv}^{\rm vib} + \tE_{J,k}^{\varv,\rm rot}.
\end{equation}
Additionally, when generating molecular cross sections, ExoCross can handle the values $F_{\varv}$  as external parameters \citep{21ClYu}.

Expanding this to include both vibrational and rotational temperatures, in this work we use a simplified bi-temperature Treanor approximation model and investigate spectral signatures of different vibrational bands contributing to a generic non-LTE spectrum.   
In the Treanor approximation, the vibrational populations are modelled  by a vibrational Boltzmann distribution but for a different temperature $T_{\rm vib}$, with the total non-LTE population of a given state taken as the product of the two Boltzmann distributions: 
\begin{equation}
\label{e:F:non-LTE}
F_{J,\varv,k}(T_{\rm vib},T_{\rm rot}) = \frac{ g_{J,k,\varv}^{\rm tot} \, e^{-c_2 \tE_{\varv}^{\rm vib}/T_{\rm vib}} e^{-c_2 \tE_{J,k}^{\varv,\rm rot}/T_{\rm rot}}}{Q(T)}. 
\end{equation}
where  $Q(T)$ is the non-LTE partition function defined as a sum over states:
\begin{equation}
\label{e:pf:2}
  Q(T) =\sum_{n}  g_{J,k,\varv}^{\rm tot} e^{-c_2 \tE_{\varv}^{\rm vib}/T_{\rm vib}} e^{-c_2 \tE_{J,k}^{\varv,\rm rot}/T_{\rm rot}}. 
\end{equation}

In ExoCross, the vibrational part  $\tE_{\varv}^{\rm vib}$ in Eq.~\eqref{e:Erot:Evib}  is taken as the lowest energy for a given $\varv$ (e.g. for $J=0$), while the rotational contribution is then estimated via Eq.~(\ref{e:Erot:Evib}) as a difference 
\begin{equation}\label{e:Erot:Evib:J:k}
  \tE_{J,k}^{\varv,\rm rot} = \tE_{\varv,J,k} - \tE_{\varv}^{\rm vib}.
\end{equation}

An absorption line intensity $ I({\rm f} \gets {\rm i})$ (cm/molecule) is then given by
\begin{equation}
\label{e:int}
 I({\rm f} \gets {\rm i}) = \frac{ A_{\rm fi}}{8 \pi c \tnu_{\rm fi}^2}  F_{J,\varv,k}(T_{\rm vib},T_{\rm rot}) \left( 1-e^{-c_2\tilde{\nu}_{\rm fi}/T} \right),
\end{equation}
where $A_{\rm fi}$ is the Einstein-A coefficient ($s^{-1}$), $\tilde{\nu}_{\rm fi}$ is the transition wavenumber (\cm).

As discussed in section \ref{sec:intro}, the bi-temperature parametrization is commonly used in laboratory quantifications of non-LTE conditions. It also introduces a simplistic `non-LTE parameter' ($T_{\rm vib}$) that could be useful for atmospheric retrieval in the exoplanet case, which in a more realistic non-LTE model can be extended by introducing different $T_{\rm vib}$ for individual vibrational states or groups of states\citep{01LoTaxx}. In the case of highly symmetric polyatomic molecules such as CH\4, such additional vibrational temperatures may be necessary to adequately describe the spectrum. For \mbox{\citep{20DuSuBr.CO}}, a tri-temperature parameterisation was necessary for CH\4: two vibrational temperatures were employed to account for differences between the stretching and bending modes of the molecule. Although this potential limitation should be borne in mind, it is worth noting that the bi-temperature model can provide sufficient descriptive performance for CH\4 in some conditions \mbox{\citep{20BuStBe.CH4}}, and remains a simple method for introducing non-LTE effects where more complicated models are not possible or prohibitively difficult to compute. In addition, it can prove particularly appropriate when focusing on limited wavelength ranges, within which a particular mode dominates, reducing the need for additional vibrational temperatures.

Regardless of the model used, the non-LTE effects are largely affected by the vibrational populations of the individual states, which affects the shape of the spectrum by modifying  intensities of the corresponding absorption bands. It is therefore important to characterize the individual contributions as well as their sensitivity to the non-LTE population input; here we make use of Exomol state assignments to plot these indvidual contributions.
Armed with cross sections for molecules in LTE and non-LTE, our goal is thus to evaluate distinguishability of specific vibrational bands by evaluating the differences between the two cases; from this we can narrow down the wavelength ranges at which the differences are most apparent and with what amplitudes.

Two planets were chosen to provide reference values as case studies with which to contrast the LTE and non-LTE cases in molecular cross sections. 
The WASP-12b reference case for modelling presents a target that has been well studied in the literature and is a perfect example of the extreme physical conditions which lend themselves to sustained non-LTE effects. These extreme conditions present in the atmosphere of WASP-12b include strong thermal heating and the appearance of high atmospheric inflation \citep{13SwDeTi.exo} these conditions are coupled with, and in part driven by, the planet's close proximity to its host star. This closeness also places WASP-12b in prime position for strong non-LTE conditions to arise in its atmosphere on account of stellar irradiation. The extreme conditions do not end there however, there is evidence to suggest that WASP-12b is also undergoing mass loss \citep{19BeZhCu} and even that its orbit is decaying \citep{19YeWiKn}.

WASP-76b's atmosphere was used as a template for generating TiO cross sections due to the detection of TiO in its atmosphere \citep{20EdChBa.exo}.

In principle it is reasonable to associate the rotational temperature  to the  equilibrium (kinetic) temperature. However one can also argue that the presence of non-LTE in the spectrum of an exoplanet can affect the equilibrium temperature  if the retrieval model is based on the assumption of pure LTE, especially if the spectral range is very limited or due to any other shortcomings. We therefore consider two temperature scenarios when considering non-LTE spectral features. For scenario~1 the planet atmospheric  equilibrium (reference) temperature (of 1864~K \ref{WASP12bParamsTable} in the case of WASP-12b) was taken as the vibrational temperature ($T_{\rm vib}$), while the rotational temperature ($T_{\rm rot}$) was reduced to 700~K to introduce a non-LTE temperature contrast between the two  temperatures. For scenario~2, the equilibrium atmospheric temperature was taken as the rotational temperature (a common assumption in non-LTE atmospheric studies)  on account of the fast equilibration times of the rotational degrees of freedom. The vibrational temperature is increased symmetrically with the decrease in scenario~1 (such that it is 3028~K in scenario~2, in line with day-side temperatures retrieved in the literature \citep{13SwDeTi.exo,21ArDePa}).

WASP-76b is also considered for scenario 1 in order to examine cross sections for TiO (since it has been detected in the literature), with a rotational temperature of 1800~K taken along with the equilibrium temperature of 2231~K from table \ref{WASP12bParamsTable} as the vibrational temperature. This gradient of 431~K is conservative when compared to day-night side differences found from phase curves in the literature \citep{21MaKoSt.exo}, correspondingly the vibrational temperature is raised to 2662~K under scenario~2.

These non-LTE  bi-temperature cross sections are then compared to baseline LTE cross sections for temperatures based on the reference exoplanet atmospheres and, in the case of H\2O in the atmosphere of WASP-12b, used for comparison with simplistic forward modelling.

For H\2O in the atmosphere of WASP-12b, we take a step beyond the comparison of the raw cross-sections to consider the non-LTE molecular opacities in the context of other effects which contribute to the spectra; this is done through a simplistic forward modelling approach. The atmospheric retrieval and forward modelling code TauREx 3.1 is used \citep{TauREx3}. It is used to produce modelled spectra for planetary atmospheres using opacities for molecules both in LTE and non-LTE for comparison; in addition to this, it includes the spectral effects of other thermodynamic, chemical and physical phenomena so that LTE versus non-LTE differences can be evaluated in context.

For the rudimentary forward modelling of WASP-12b's atmosphere, we focus on the water molecule to evaluate the differences between the non-LTE and LTE cases, owing to its observability with the Hubble Space Telescope’s WFC3 instrument. To conduct this initial forward modelling, custom cross sections \citep{jt801} for  water  were used based on the   POKAZATEL line list \citep{jt734}.

\begin{table}
	\caption{Parameters used to inform cross section calculations and in forward modelling.}
	\label{WASP12bParamsTable}
	\footnotesize
	\centering

		\begin{tabular} {c | c | c}
			\hline
			\hline
			Parameter & WASP-12b & WASP-76b \\
			\hline
			Planet Radius ($R_{J}$) & 1.9 & 1.854  \\
			Planet Mass ($M_{J}$) & 1.47 & 0.894 \\
			Planet Equilibrium Temperature (K)& 1864 & 2231\\
			Stellar Temperature (K) & 6360 & 6329\\
			Stellar Radius ($R_{\odot}$) & 1.657 & 1.756\\
			Cloud $\log{}$ Pressure (Pa) & 2.38 & 8.128\\
			H\2O $\log{}$Mixing Ratio & -3.12 & -2.85\\
			CO\2\  $\log{}$Mixing Ratio & -9.00 & - \\
			TiO  $\log{}$Mixing Ratio & - & -5.62\\
			Transit Duration (h) & 3.00 & 3.69 \\
			Stellar $\log{}g$ (cm$/$s$^{2}$)& 4.15 & 4.196\\
			Stellar Metallicity (dex) & 0.3 & 0.23\\
			Inclination &  $83.37^{\circ}$ & $89.623^{\circ}$\\
			Semi-major Axis (AU) & 3.039 &  0.0334\\
			Period (Days) & 1.091 & 1.8099\\  

			\hline
	\end{tabular}
\end{table}

\section{Case Studies}

For the molecules considered in this paper (CH\4, H\2O, CO and TiO) we primarily examine cross sections directly, using reference temperatures from WASP-12b and WASP-76b for two non-LTE scenarios. Although molecular non-LTE examples for atmospheres within our solar system are commonly seen in emission, the choice to explore these effects on absorption cross sections is motivated by demonstrating the application to exoplanet atmosphere retrieval codes. These codes ingest absorption cross section data for use in the modelling of exoplanet atmospheres. These allow us to see how individual absorption transitions originating from over- or under-populated lower vibrational states affect the  molecular absorption due to non-LTE effects. Individual vibrational bands are characterised by different shapes, which are independent from the LTE or non-LTE model. Their spectral mixture is however defined by the specific set of the (ro-)vibrational populations. These shapes are distinct even at low resolution due to their broad natures and can be used as indicators of the presence of (specific) non-LTE effects. 
In this work the individual vibrational bands are generated as absorption cross sections with the help of  the vibrational quantum numbers provided in the ExoMol line lists. The bi-temperature non-LTE models with the high-contrast temperatures described above present a convenient tool to visualise the signatures of possible non-LTE vibrational effects.

Examining the raw molecular cross sections and identifying the distinctions between the LTE and non-LTE cases are necessary to identify non-LTE ‘signatures’. These signatures represent the ideal scenario of distinguishing non-LTE from LTE at the raw cross section level; this is before additional atmospheric factors are accounted for by forward modelling, and without considering instrument limitations. In some cases large differences in the spectra are observed that may be visible at lower resolution and in some cases there are also many individual lines that differ between the LTE and non-LTE spectra; such differences in the individual lines may be resolvable with high resolution spectroscopy performed from the ground. In these cross sections, we can first identify non-LTE signatures where the spectrum for a given molecule in non-LTE differs from that of the same molecule in LTE. 

\subsection{Absorption Water Cross Sections}

The difference in the non-LTE and LTE spectra of water can be illustrated by analysing intensities of the individual vibrational bands. Indeed, according to  Eq.~(\ref{e:F:non-LTE}) and also assuming that the non-LTE rotational temperature matches the LTE equilibrium temperature (scenario 2: $T_{\rm rot} = T_{\rm eq}$), a non-LTE spectrum can be understood as a combination of vibrational bands that correspond to higher vibrational temperatures. Under the LTE conditions, only the lowest vibrational states are significantly populated, with the  vibrational ground state (g.s.) the most-populated, and, depending on  energies and temperature, the first excited vibrational states. In the case of the H\2O molecule these are $\nu_1$ (symmetric stretch), $\nu_2$ (symmetric bend) and $\nu_3$ (asymmetric stretch), where the energy of $\nu_2$ is almost half of $\nu_1$ or $\nu_3$. A non-LTE spectrum with a higher vibrational temperature ($T_{\rm vib}> T_{\rm rot}$) contains transitions corresponding to excited vibrational states beyond the ground vibrational state. 

\begin{figure*}
\includegraphics[width=0.48\textwidth]{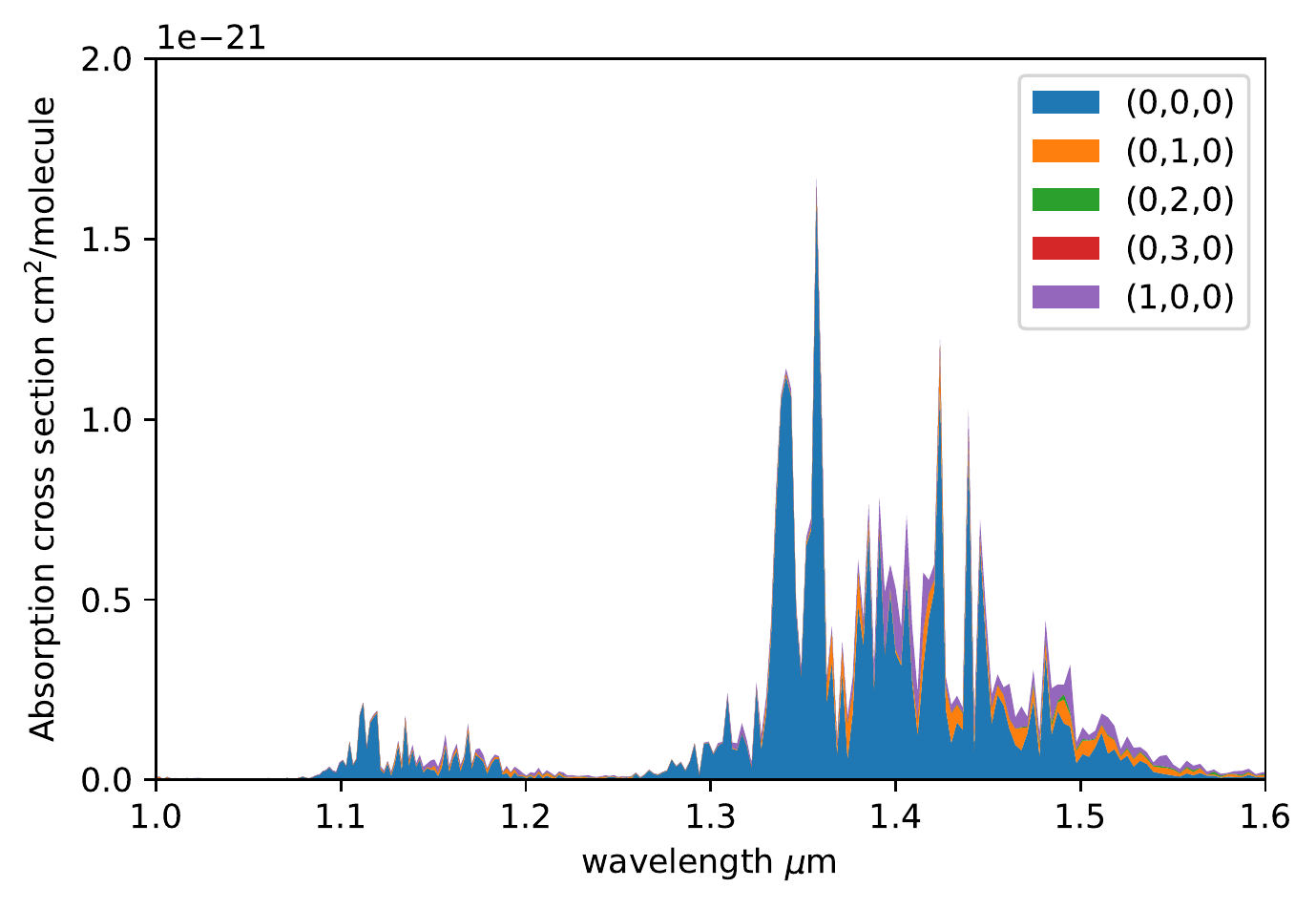}
\includegraphics[width=0.48\textwidth]{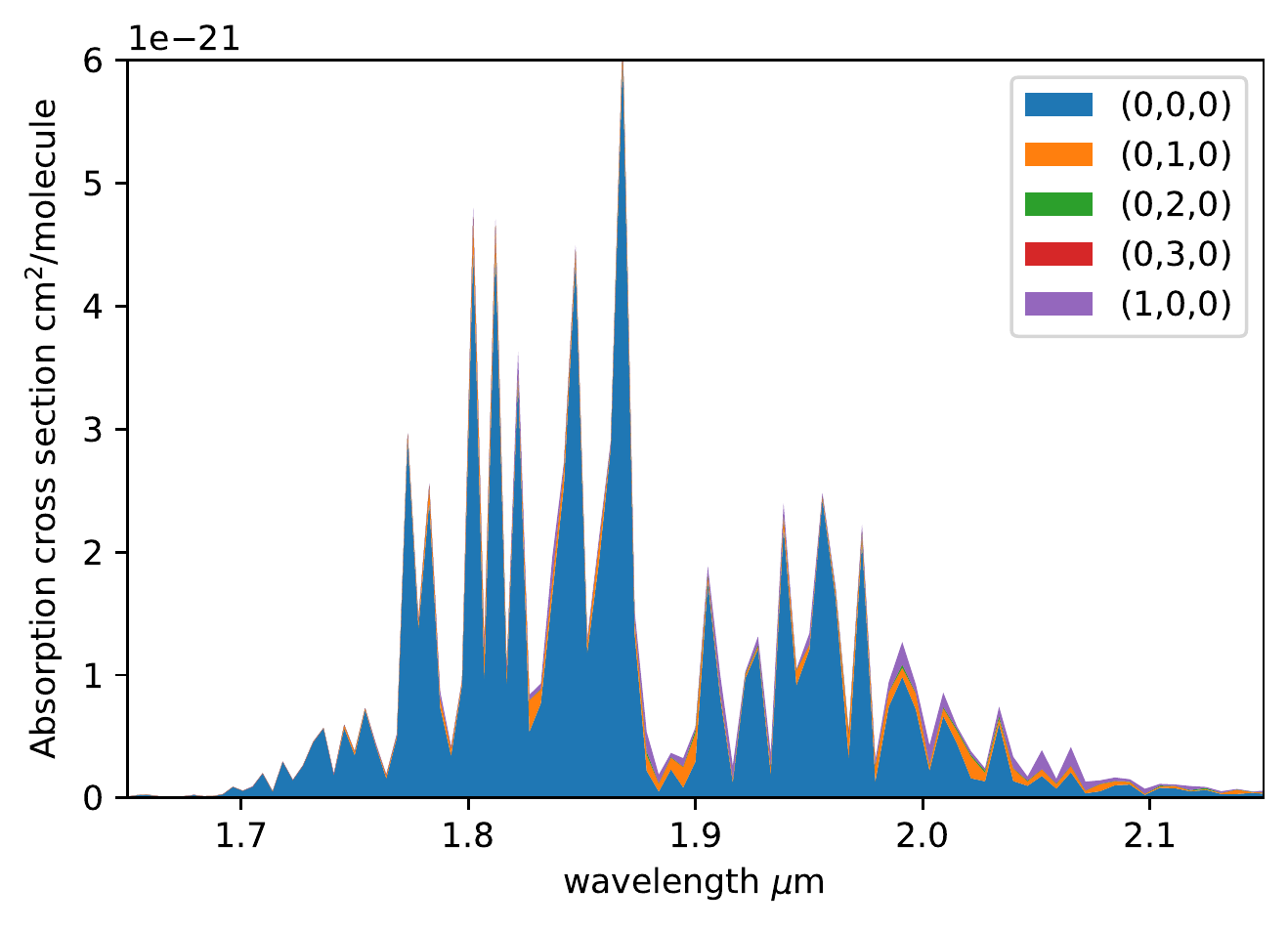} \\
\includegraphics[width=0.49\textwidth]{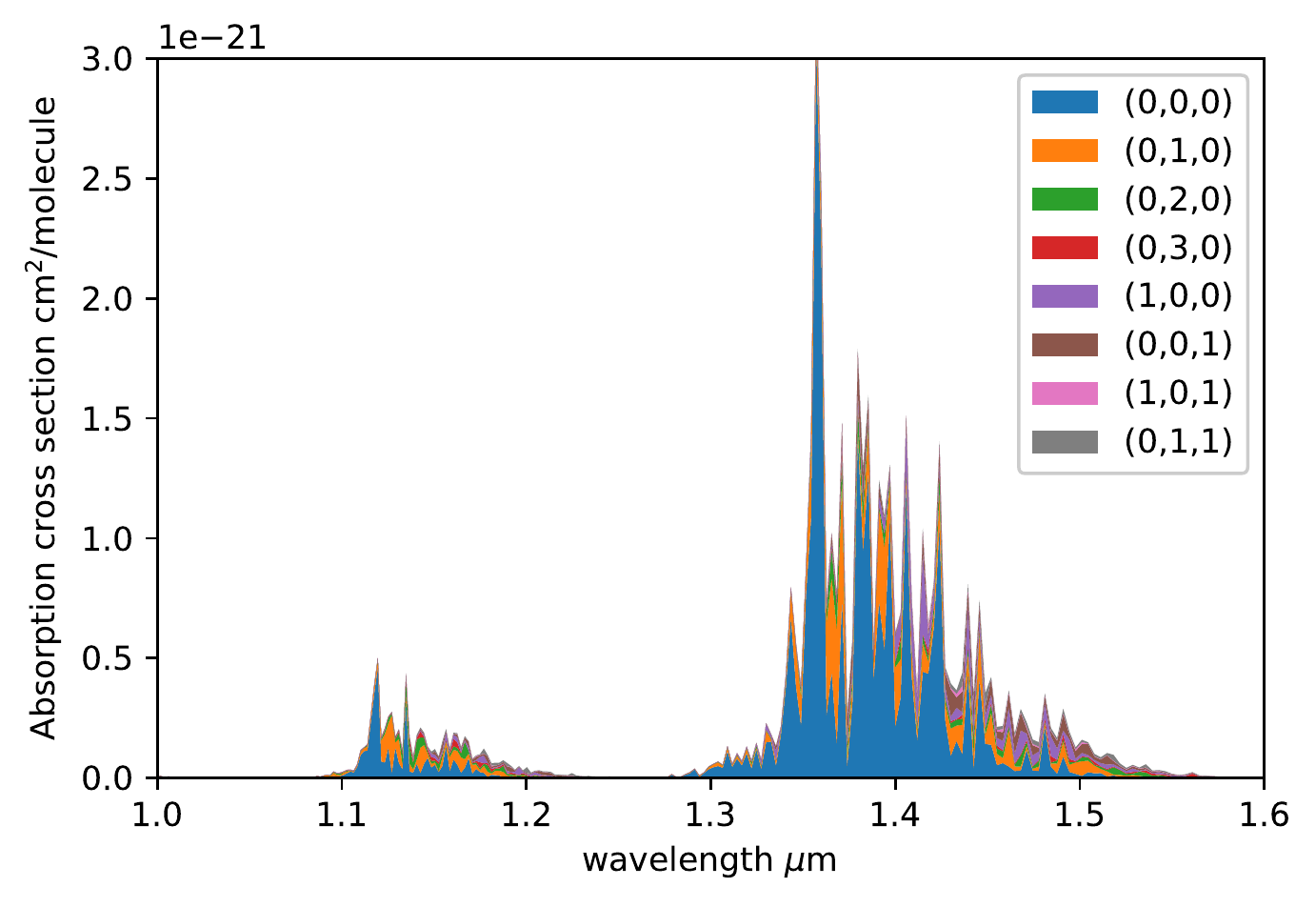}
\includegraphics[width=0.49\textwidth]{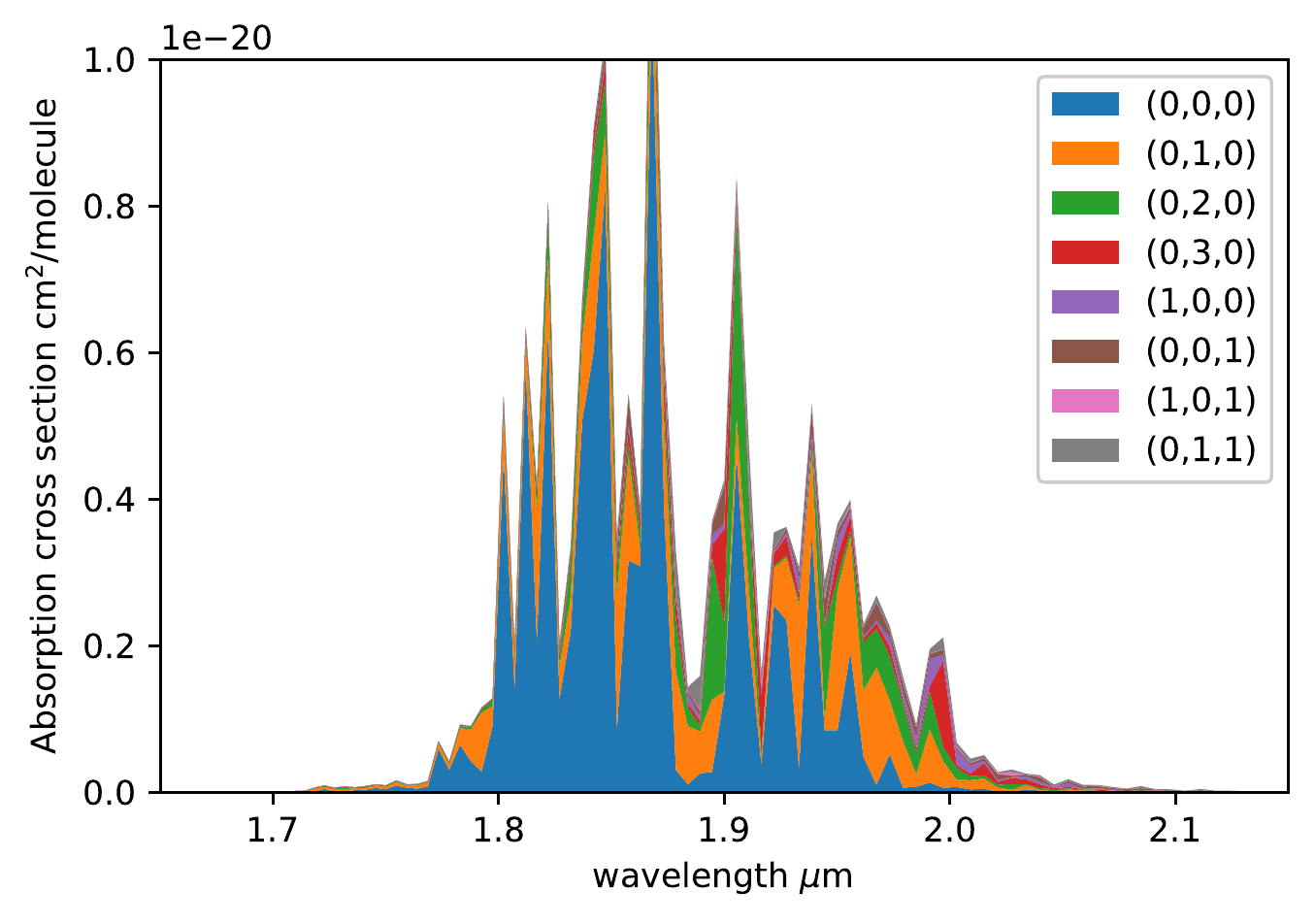} \\
\includegraphics[width=0.49\textwidth]{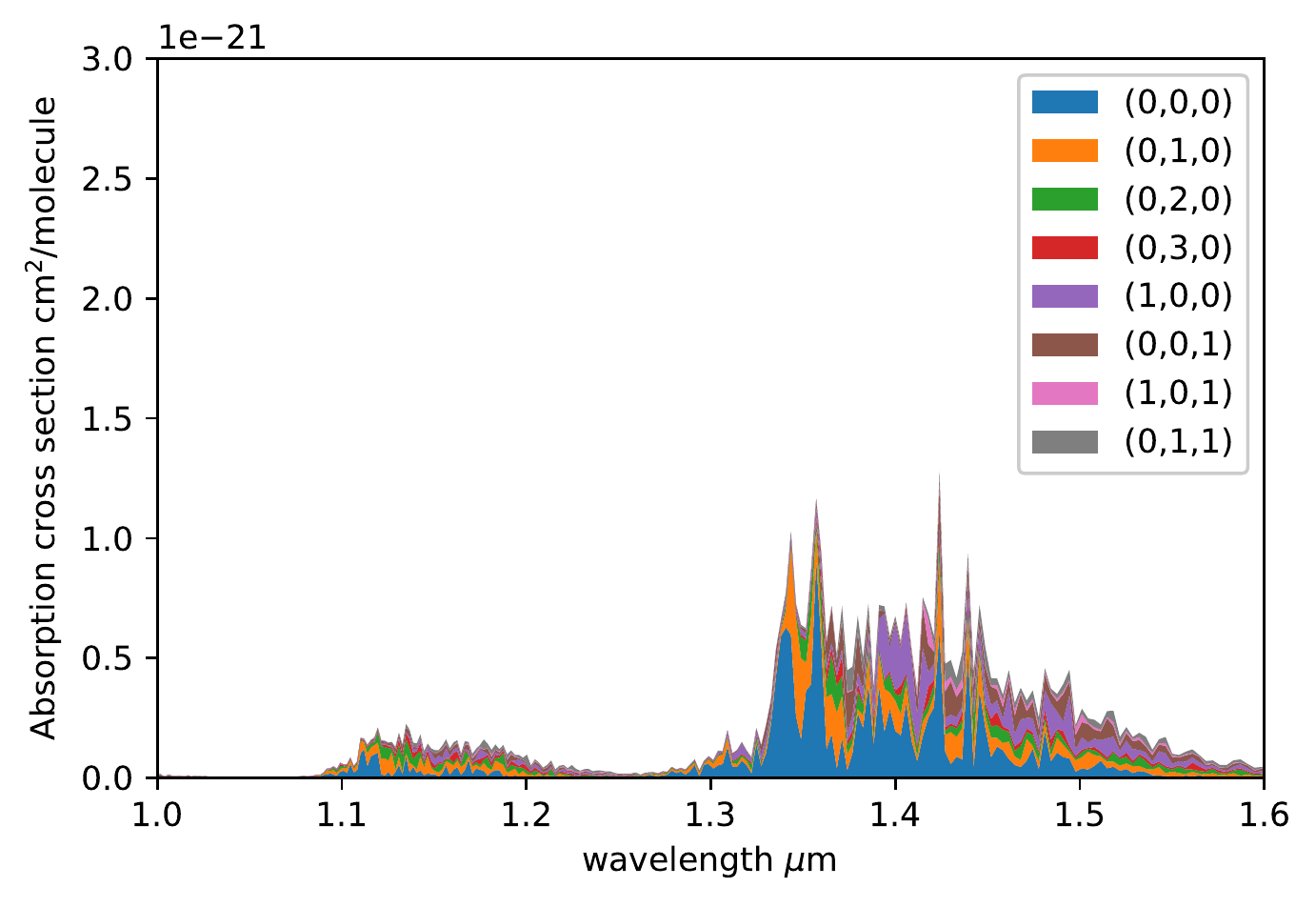}
\includegraphics[width=0.49\textwidth]{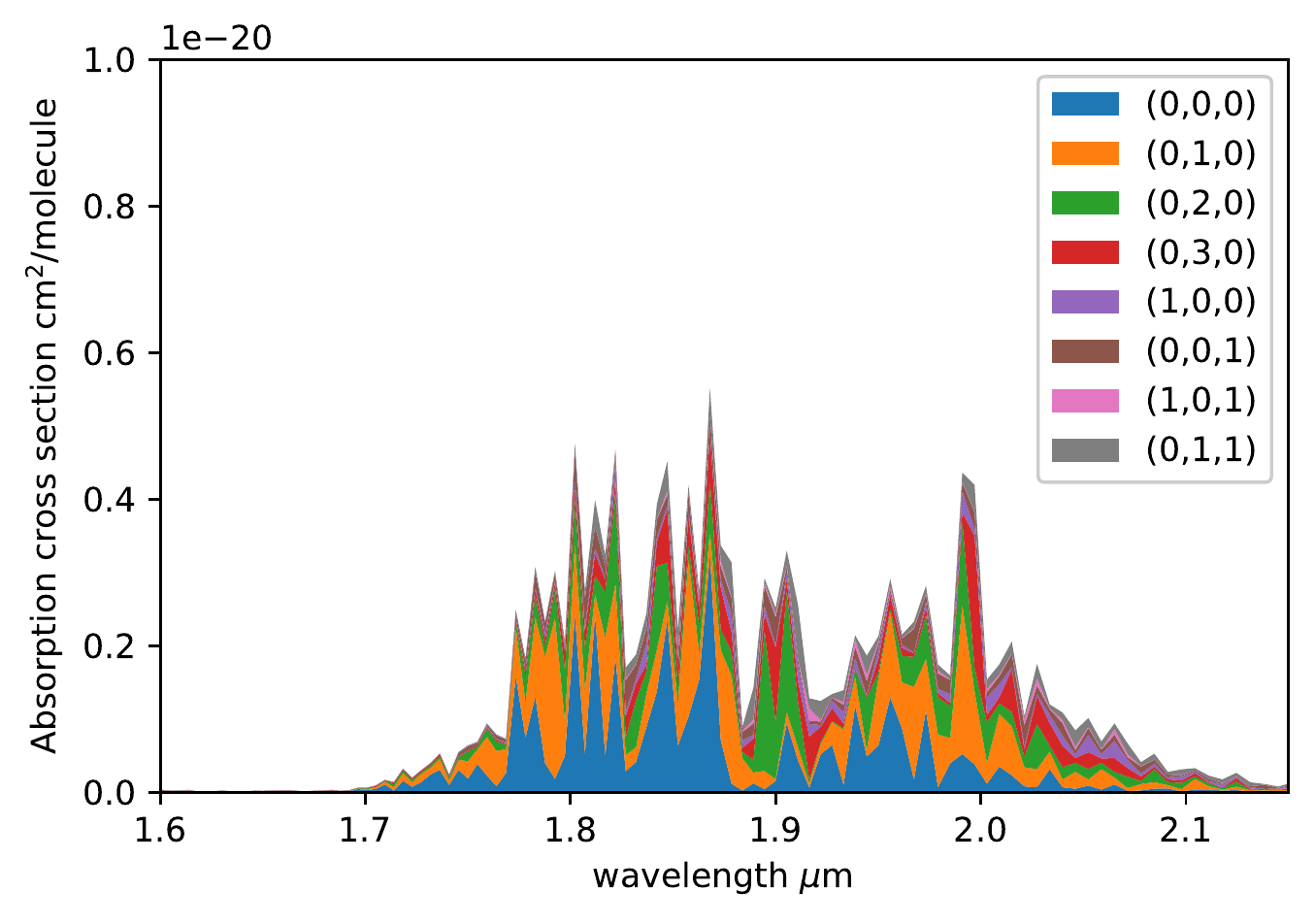}
\caption{ \label{fig:H2O_bands}
The individual contributions to the
$2\nu_2$ (left column)  and $\nu_1+\nu_2$ (right column) systems from different
absorption bands for H\2O at 1 bar pressure. Shown for: LTE ($T = 1864$~K) (top row), non-LTE scenario~1 ($T_{\rm rot}=$ 700~K, $T_{\rm vib}$ = 1864~K) (middle row) and
non-LTE scenario~2 ($T_{\rm rot}= $ 1864~K,  $T_{\rm vib}$ = 3028~K) (bottom row). Legends indicate  quantum numbers $(v_1,v_2,v_3)$ of the lower vibrational states. }
\end{figure*}

Figure~\ref{fig:H2O_bands} shows  cross-section contributions from several   vibrational bands of H$_2$O plotted on top of each other,  both for the LTE and non-LTE absorption spectra, where the latter are given for both scenario~1 and 2. The LTE spectrum was computed using $T$ = 1864~K, while the non-LTE spectra correspond to $T_{\rm rot} = 700$~K and $T_{\rm vib} = 1864$~K for scenario~1 and $T_{\rm rot} = 1864$~K and $T_{\rm vib} = 3028$~K for scenario~2, all with pressure broadening consistent with 1 bar atmospheric pressure. Different colours represent  bands originating from different vibrational states $(\varv_1'',\varv_2'',\varv_3'')$, where $\varv_1''$, $\varv_2''$ and $\varv_3''$ are the standard normal more quantum numbers  corresponding to the (lower state) symmetric stretch, symmetric bend and asymmetric stretch, respectively.  For example, $(1,0,0)\gets (0,0,0)$ with $(\varv_1'',\varv_2'',\varv_3'')=(0,0,0)$ corresponding to the fundamental band $\nu_1$, which originates in  the ground vibrational state, while  $(1,1,0)\gets (0,1,0)$ is an example of a hot band  originating in the vibrationally excited state $(0,1,0)$. Only quantum numbers of the lower states  $(\varv_1'',\varv_2'',\varv_3'')$ are indicated in this figure. 

As expected, the LTE spectrum of H$_2$O at $T=1864$~K in figure~\ref{fig:H2O_bands} is dominated by the overtone bands starting from the ground vibrational state  $(0,0,0)$ with some contribution from the hot band originating from the bending state ($0,1,0)$. The latter has a similar profile and therefore does not affect the overall shape of the spectrum. 

In the case of the $T_{\rm rot}= $ 700~K, $T_{\rm vib} = 1864$~K non-LTE simulations (scenario 1),  the contributions from the hotter bands become significant. What is important here is that the stretching excitations are characterized by different profiles. As a result, the  change becomes significant even for lower resolutions. For example, strong structures at $1.4$~$\mu$m or $1.95$~$\mu$m come  from the hot bands  solely due to the non-LTE vibrational excitations of $T_{\rm vib} = 1864$~K. The contribution to the signature non-LTE shoulder can be seen most distinctly in figure \ref{fig:H2O_bands} where the excited bands cause substantial divergence from the LTE case between 1.9 and 2.05~$\mu$m.

When considering the second non-LTE scenario, i.e. for  $T_{\rm rot}= $ 1864~K,  $T_{\rm vib} = 3028$~K, the differences become even more apparent. Comparing the bottom panel of figure \ref{fig:H2O_bands} to the middle panel, the additional contributions to the spectra of states other than the ground vibrational state $(0,0,0)$ are shown to be in excess for scenario~2.

\begin{figure}
\centering
\includegraphics[width=\columnwidth]{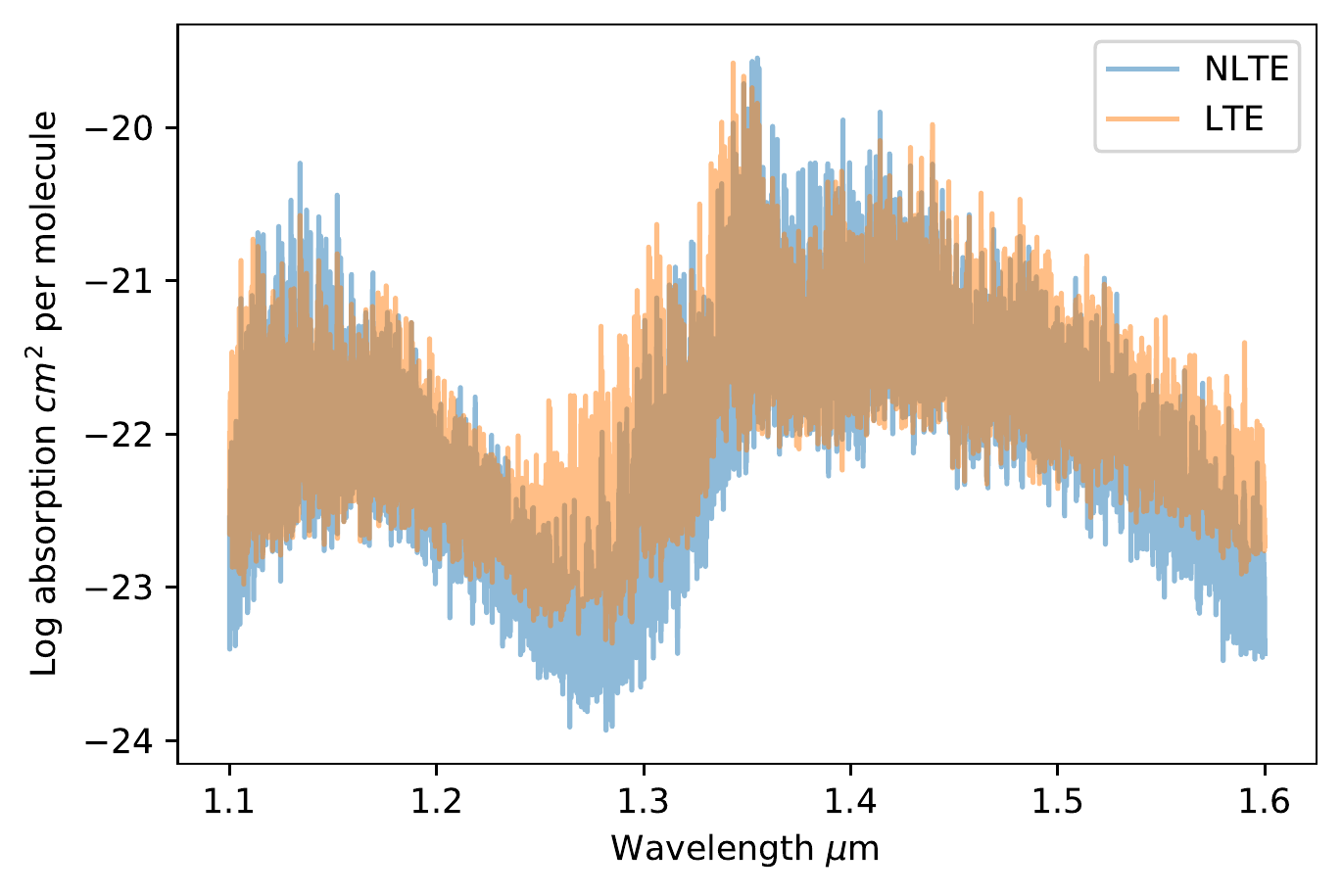}\\
\includegraphics[width=\columnwidth]{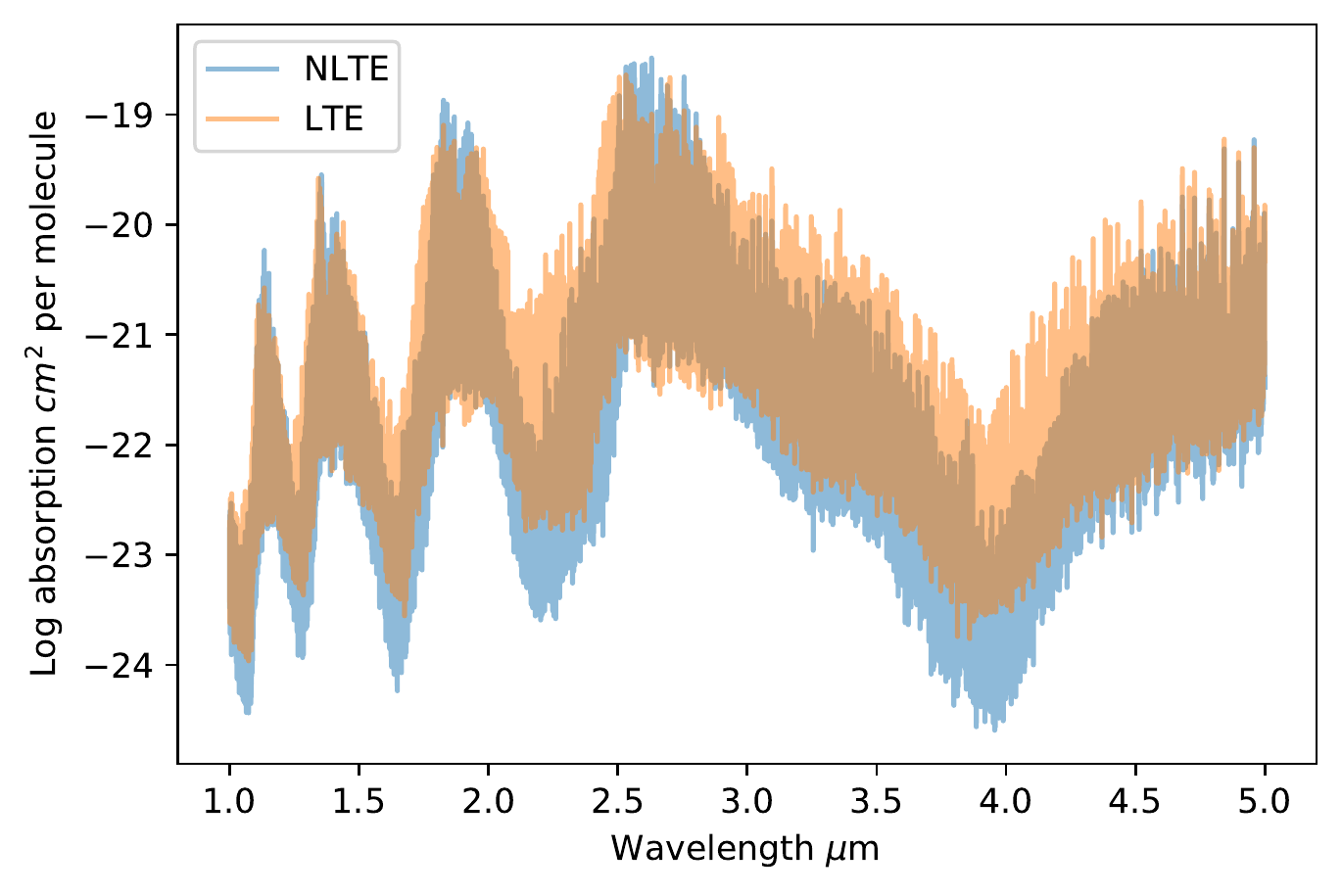}
\caption{Opacity cross sections for H\2O in LTE and non-LTE at WASP-12b's equilibrium temperature (1864~K) and a pressure of 1 bar, with a rotational temperature of 700~K for the non-LTE case in the 1.1 to 1.6 \um\ region (Top display) and   1 to 5 \um\ region (Bottom display). \label{fig:H2O_xsec_scenario_1}}
\end{figure}

\begin{figure}
\centering
\includegraphics[width=\columnwidth]{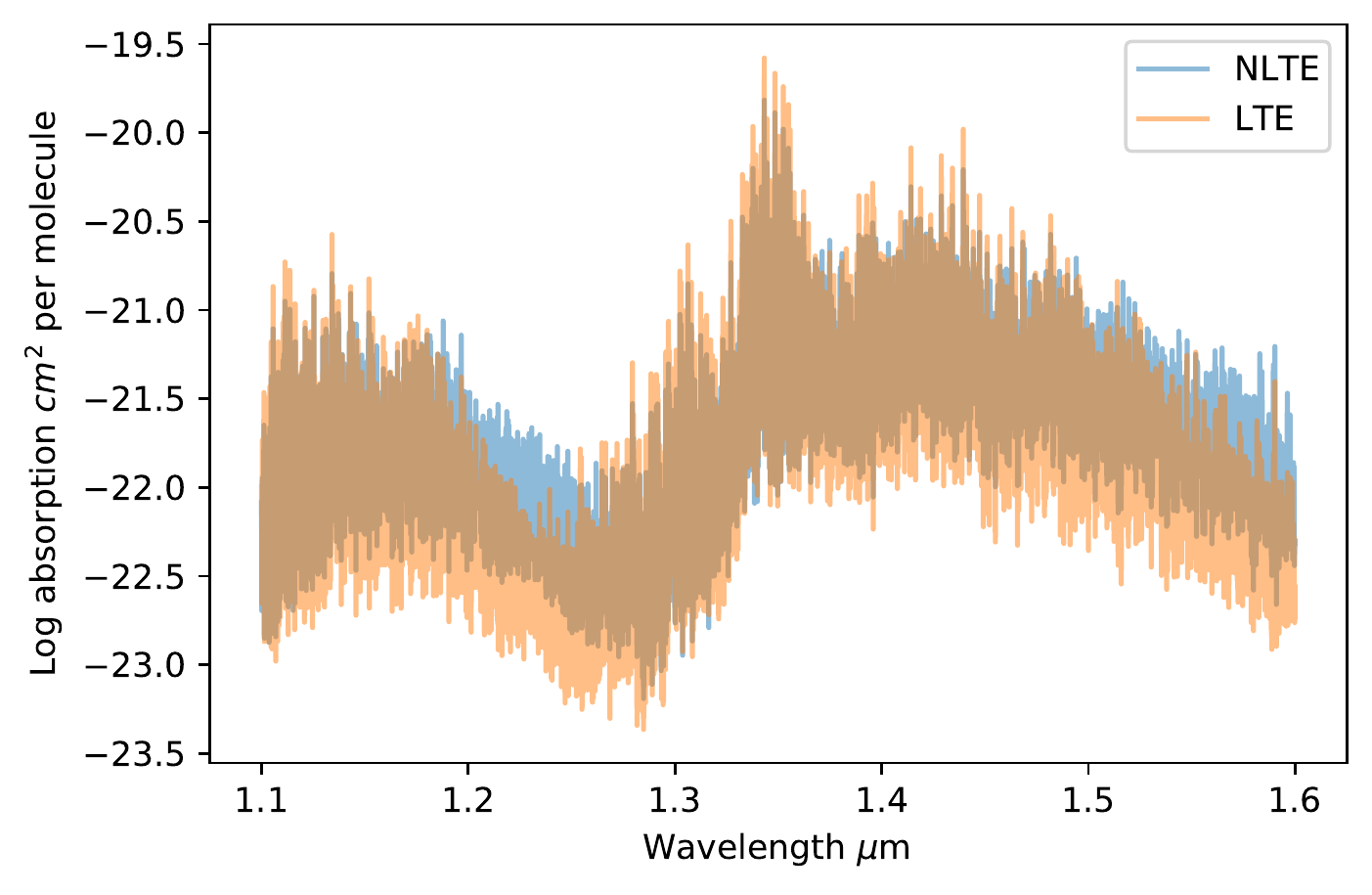}\\
\includegraphics[width=\columnwidth]{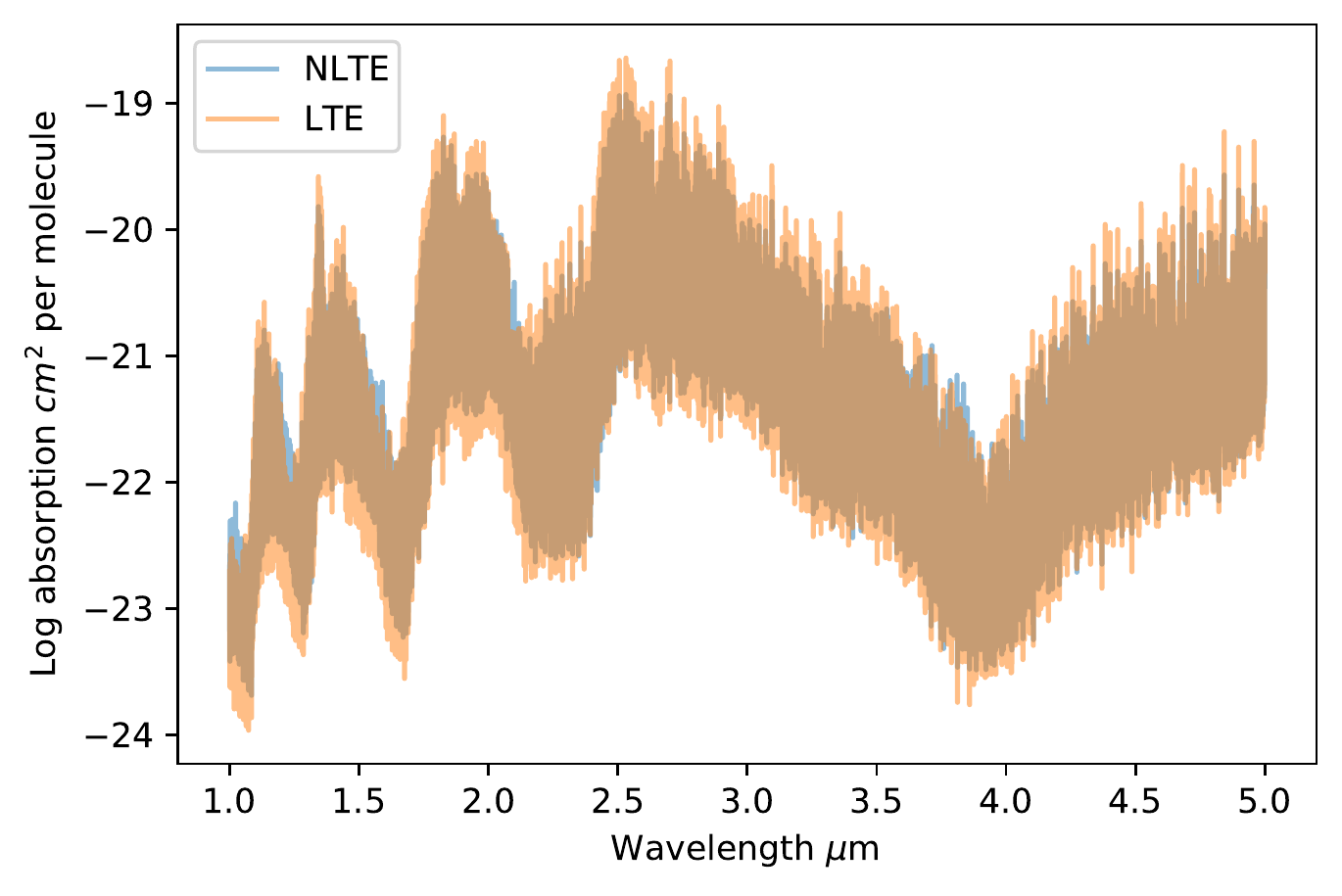}
\caption{Opacity cross sections for H\2O in LTE and non-LTE at WASP-12b's equilibrium temperature (1864~K) and a pressure of 1 bar, with a vibrational temperature of 3028~K for the non-LTE case in the 1.1 to 1.6 \um\ region (Top display) and   1 to 5 \um\ region (Bottom display). \label{fig:H2O_xsec_scenario_2}}
\end{figure}

Figure~\ref{fig:H2O_xsec_scenario_1}  compares  opacity cross sections for H\2O in LTE and non-LTE at WASP-12b's equilibrium temperature (1864~K) and a pressure of 1 bar, with a rotational temperature of 700~K for the non-LTE case assuming scenario~1 (reduced $T_{\rm rot}$). This is a large temperature contrast case chosen to demonstrate non-LTE effects that can arise in exoplanet atmospheres. 
For H\2O, as shown in figure \ref{fig:H2O_xsec_scenario_1}, the characteristic difference between the non-LTE and LTE cases are  `shoulders' present where H\2O is in LTE, but absent when in non-LTE, which are also shown in figure \ref{fig:H2O_xsec_scenario_1} (left display) as a zoom-in of the 1.4\,$\mu$m band.  These shoulders are formed from the  rotational structure of a given vibrational fundamental band i.e. having originated in the ground vibrational state. They are predominantly from the $(0,1,0)$, $(0,2,0)$ and $(0,3,0)$ bands and are thus characteristic for the rotational temperature.  The narrowing of the  shoulders in the case of a lower rotational temperature  can be significant and thus represents the characteristic non-LTE signature. The detection potential is greatest in the infrared region of the electromagnetic spectrum; at wavelengths below this - in the optical - absorption is too low to be visible above other atmospheric effects, such as Rayleigh scattering. In addition to sufficient absorption by the molecule in the IR, the greatest departures from the LTE case occur here. LTE and non-LTE cross sections are contrasted for the second non-LTE scenario in figure\,\ref{fig:H2O_xsec_scenario_2}; under this scenario, distinctions between the two cases remain although they are less pronounced and so to distinguish the two cases requires focusing on the bands below 2~\um.

\begin{figure} 
\centering
\includegraphics[width=\columnwidth]{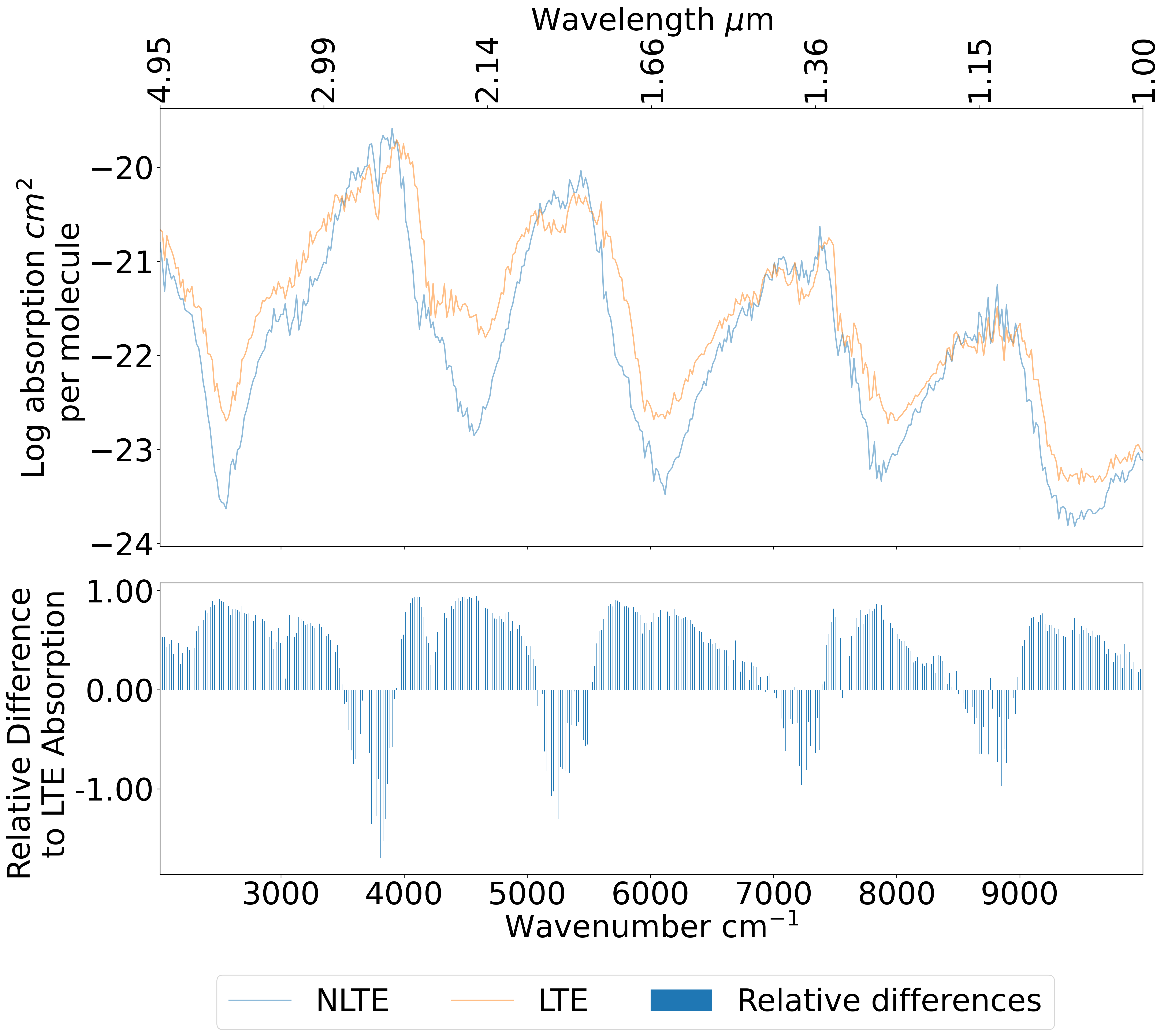}
\caption{Opacity cross sections in the 1 to 5 \um\ region for H\2O in LTE and non-LTE at WASP-12b's equilibrium temperature (1864~K) and a pressure of 1 bar, with a rotational temperature of 700~K for the non-LTE case. Down-binned to the resolving power of R $\sim$ 150 with the difference in absorption relative to the LTE case plotted beneath. \label{fig:H2O_xsec_1_5_diffs_scenario_1}}
\end{figure}

\begin{figure} 
\centering
\includegraphics[width=\columnwidth]{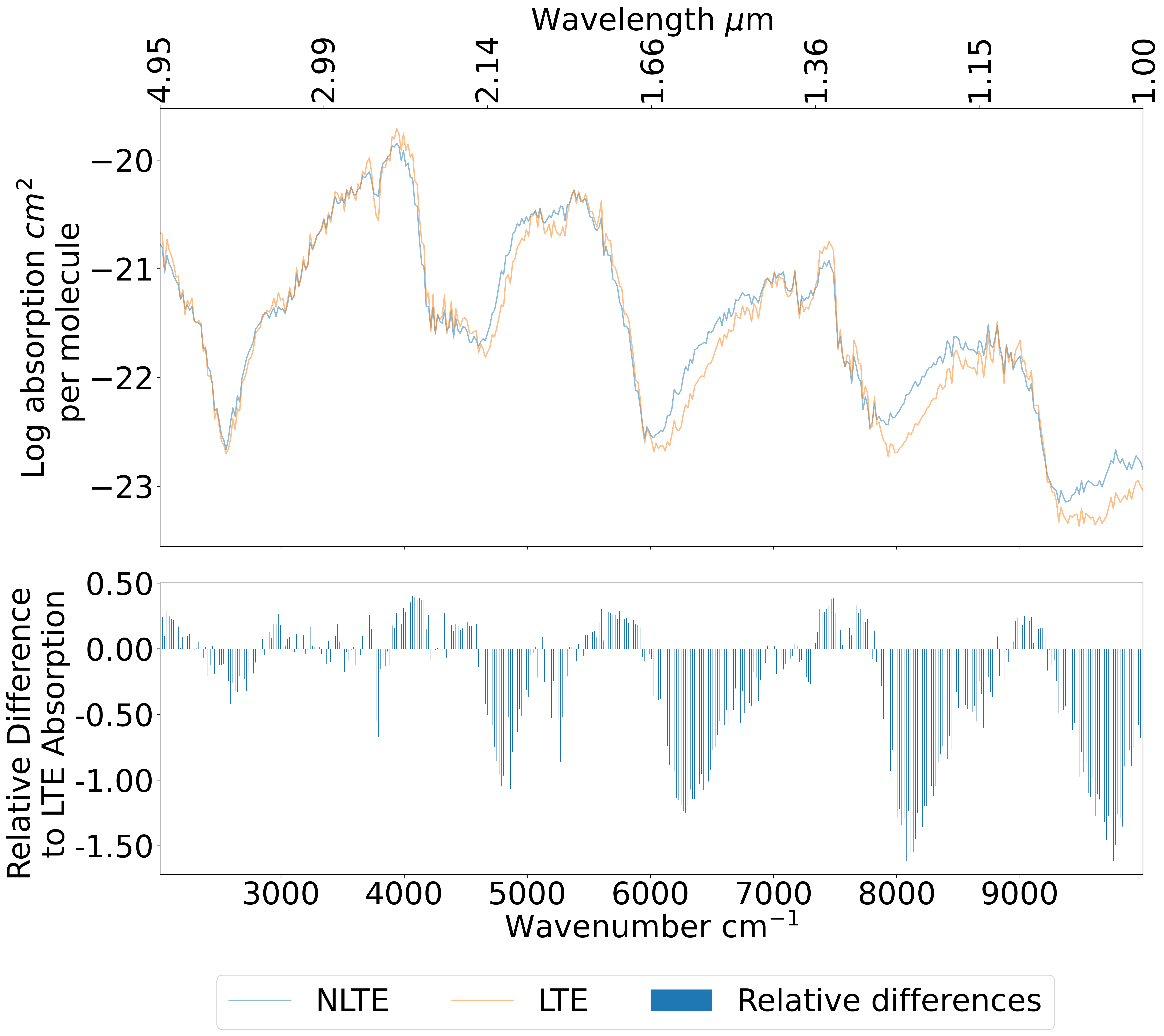}
\caption{Opacity cross sections in the 1 to 5 \um\ region for H\2O in LTE and non-LTE at WASP-12b's equilibrium temperature (1864~K) and a pressure of 1 bar, with a vibrational temperature of 3028~K for the non-LTE case. Down-binned to the resolving power of R $\sim$ 150 with the difference in absorption relative to the LTE case plotted beneath. \label{fig:H2O_xsec_1_5_diffs_scenario_2}}
\end{figure}

Figure\,\ref{fig:H2O_xsec_1_5_diffs_scenario_1} focuses on the 1 - 5\,\um\ range for scenario~1 which encompasses multiple H$_{2}$O absorption bands used in constraining the abundance of water in exoplanet atmospheres. In this wavelength range the shoulders that form the divergence between the LTE and non-LTE cases can be clearly seen when viewing both cases over-plotted, as shown at R $\sim$ 15000 in figure \ref{fig:H2O_xsec_scenario_1} and down-binned to R $\sim$ 150 in figure \ref{fig:H2O_xsec_1_5_diffs_scenario_1}. The relative differences plotted in the lower segment of figure \ref{fig:H2O_xsec_1_5_diffs_scenario_1} show absorption diverging from the LTE case by as much as 1.74x at the band around 3~\um\, while for scenario 2 figure \ref{fig:H2O_xsec_1_5_diffs_scenario_2} shows relative differences which can exceed 1.6x, albeit with a lower overall intensity in the nearer IR.  Despite harbouring differences of a greater magnitude, the low absorption in these troughs is particularly susceptible to obstruction by cloud layers in a planet's atmosphere. The Hubble Space Telescope is predisposed to the detection of water owing to the wavelength coverage of its Wide Field Camera 3; in considering its potential for observing a distinct non-LTE signature this wavelength range can be seen in figures \ref{fig:H2O_xsec_scenario_1} and \ref{fig:H2O_xsec_scenario_2} at R $\sim$ 15000. The corresponding differences for H\2O under our second scenario are shown in figure\,\ref{fig:H2O_xsec_1_5_diffs_scenario_2}. Although not as pronounced, differences remain by which the non-LTE case can be distinguished from the LTE one. These are most visible at the lower wavelengths, for instance in the wavelength range between 1.1\,\um\ and 1.6\,\um\ as shown in figure\,\ref{fig:H2O_xsec_1_5_diffs_scenario_2}'s left panel, where a flip of the shoulders now gives the non-LTE case a broader profile when vibrationally excited, compared to the rotational cooling in scenario~1.

\subsection{CO Cross Sections}

\begin{figure}
\centering
\includegraphics[width=\columnwidth]{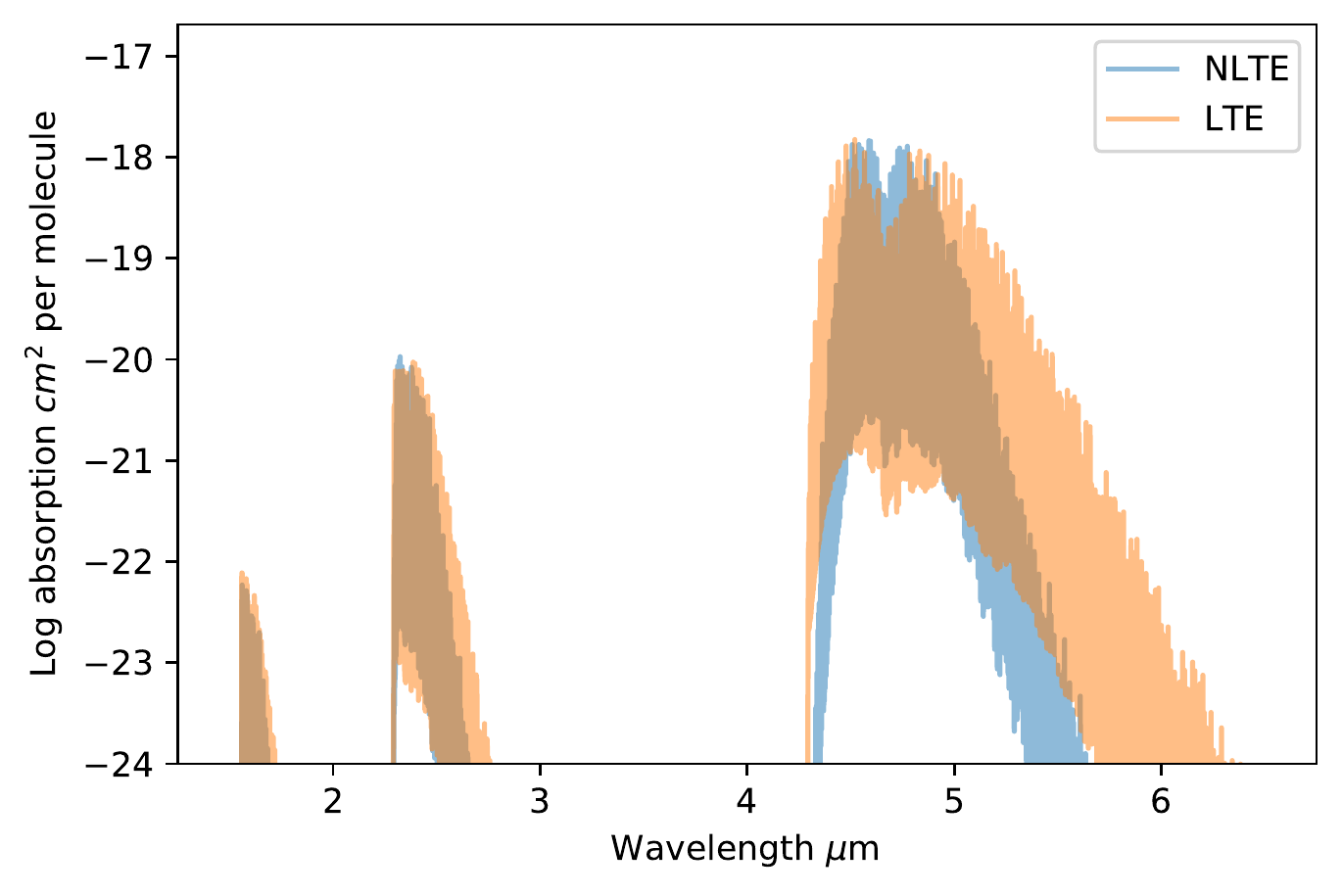}\caption{Opacity cross sections for CO in LTE (1864~K)  at WASP-12b's equilibrium temperature and and non-LTE ($T_{\rm rot}= $ 700~K, $T_{\rm vib} = 1864$~K)  for  scenario 1, with  pressure of 1 bar at R $\sim$ 15000. \label{fig:CO_Xsec} }
\end{figure}

The CO cross sections are calculated from the Li linelist \citep{15LiGoRo.CO} with Voigt line profiles at a resolution of $\sim$15,000, run across a vibrational temperature grid spanning 100~K to 3400~K at increments of 100~K. The wide ranging degree to which the CO molecule absorbs, down to less than $10^{-100}$ cm$^{2}$ per molecule, leaves only a few bands visible over other contributions in atmospheric spectra. The result is that we are only interested in a few bands; all located below 7~\um\ when we constrain to a log absorption greater than -28. These are shown in figure \ref{fig:CO_Xsec}. Of these three bands the most absorbing one lies at 5~\um\, in addition to this it presents a candidate for distinguishing the LTE and non-LTE cases owing to the offset trailing edge of the band for CO in LTE.

\begin{figure}
\centering
\includegraphics[width=\columnwidth]{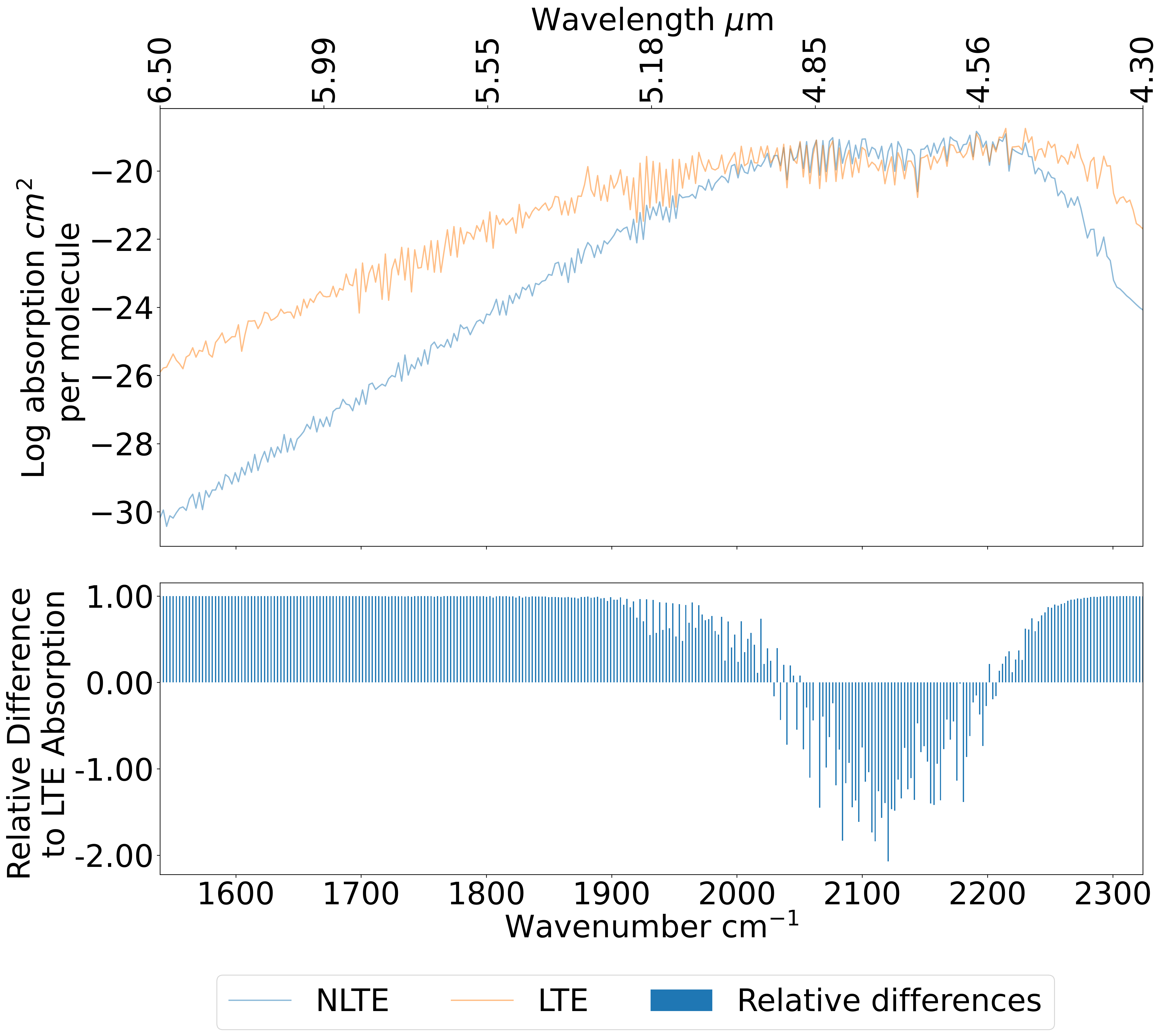}    \caption{Opacity cross sections focusing on the prominent band for CO in LTE and non-LTE at WASP-12b's equilibrium temperature (1864~K) and a pressure of 1 bar binned down to R $\sim$ 1500 for clarity, with a rotational temperature of 700~K for the non-LTE case, inline with scenario~1. \label{fig:CO_Xsec_diffs} }
\end{figure}

Using our first scenario, the non-LTE cross absorptions are generated using the bi-temperature model set with the rotational temperature as 700~K and the vibrational temperature as 1864~K, WASP-12b's equilibrium temperature.
This third band is the focus of figure \ref{fig:CO_Xsec_diffs} which shows a sizeable difference of up to 4.49x between the LTE and non-LTE cases for CO absorption in the (wavelength) band peak. In this band we can also observe a band shift where the Non-LTE intensity drops off abruptly compared to the LTE intensity, this contraction of the edge presents a good opportunity to identify Carbon Monoxide in non-LTE.

\begin{figure}
\centering
\includegraphics[width=\columnwidth]{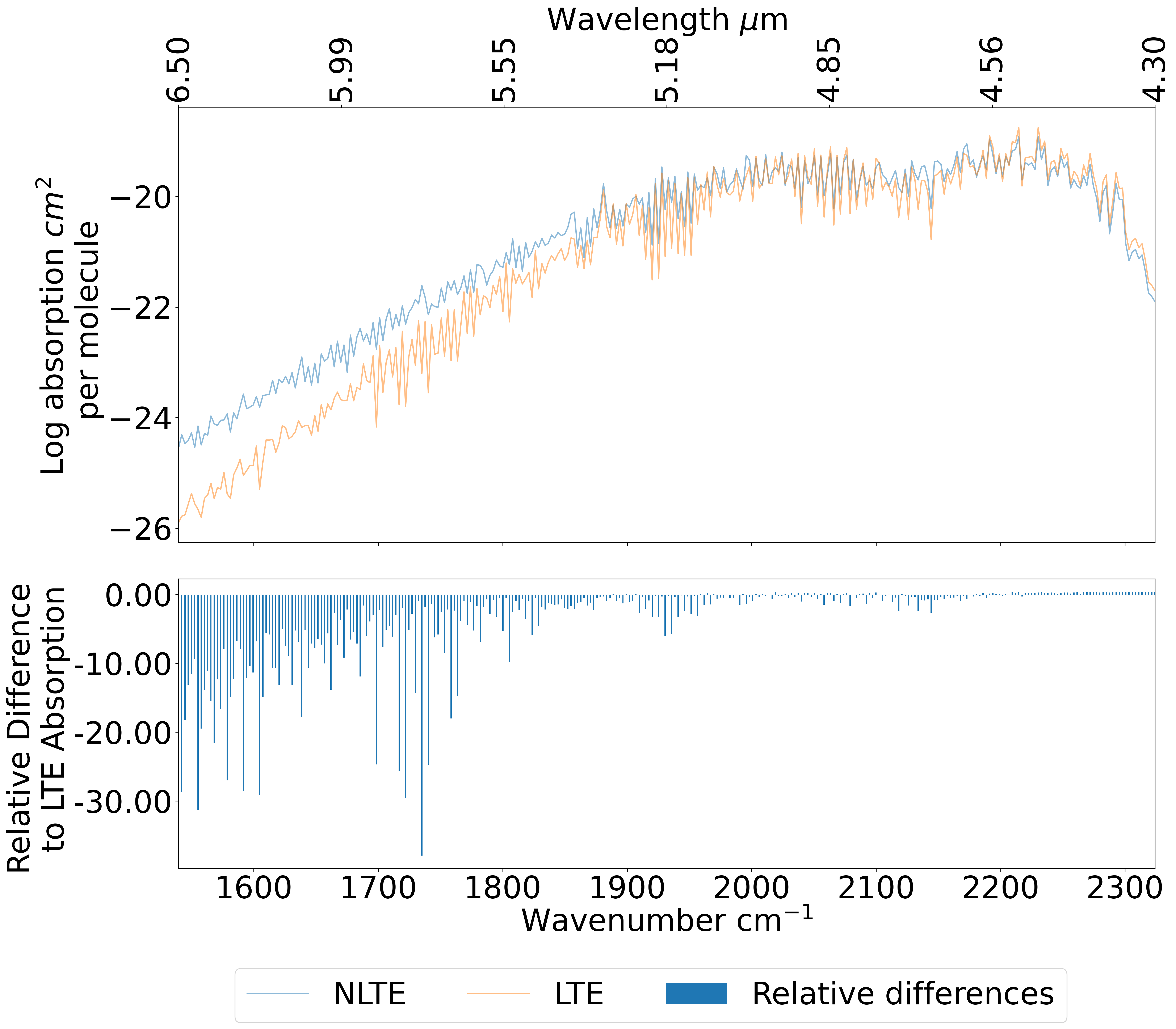}\caption{Opacity cross sections for CO in LTE and non-LTE at WASP-12b's equilibrium temperature (1864~K) and a pressure of 1 bar at R $\sim$ 1500, with a vibrational temperature of 3028~K for the non-LTE case, to demonstrate the non-LTE effect in our second scenario. \label{fig:CO_Xsec_Scenario_2} }
\end{figure}

An offset can also be seen under scenario 2 ($T_{\rm rot}= $ 1864~K,  $T_{\rm vib} = 3028$~K) in figure \ref{fig:CO_Xsec_Scenario_2}, although the difference at 6~\um ~is not as great as in scenario~1, the distinction at lower wavelength bands becomes far more useful: at 3~\um ~a larger offset emerges for scenario~2 than the much smaller offset, with a great deal of overlap, that is present in the scenario~1 case.

\begin{figure}
\centering
\includegraphics[width=\columnwidth]{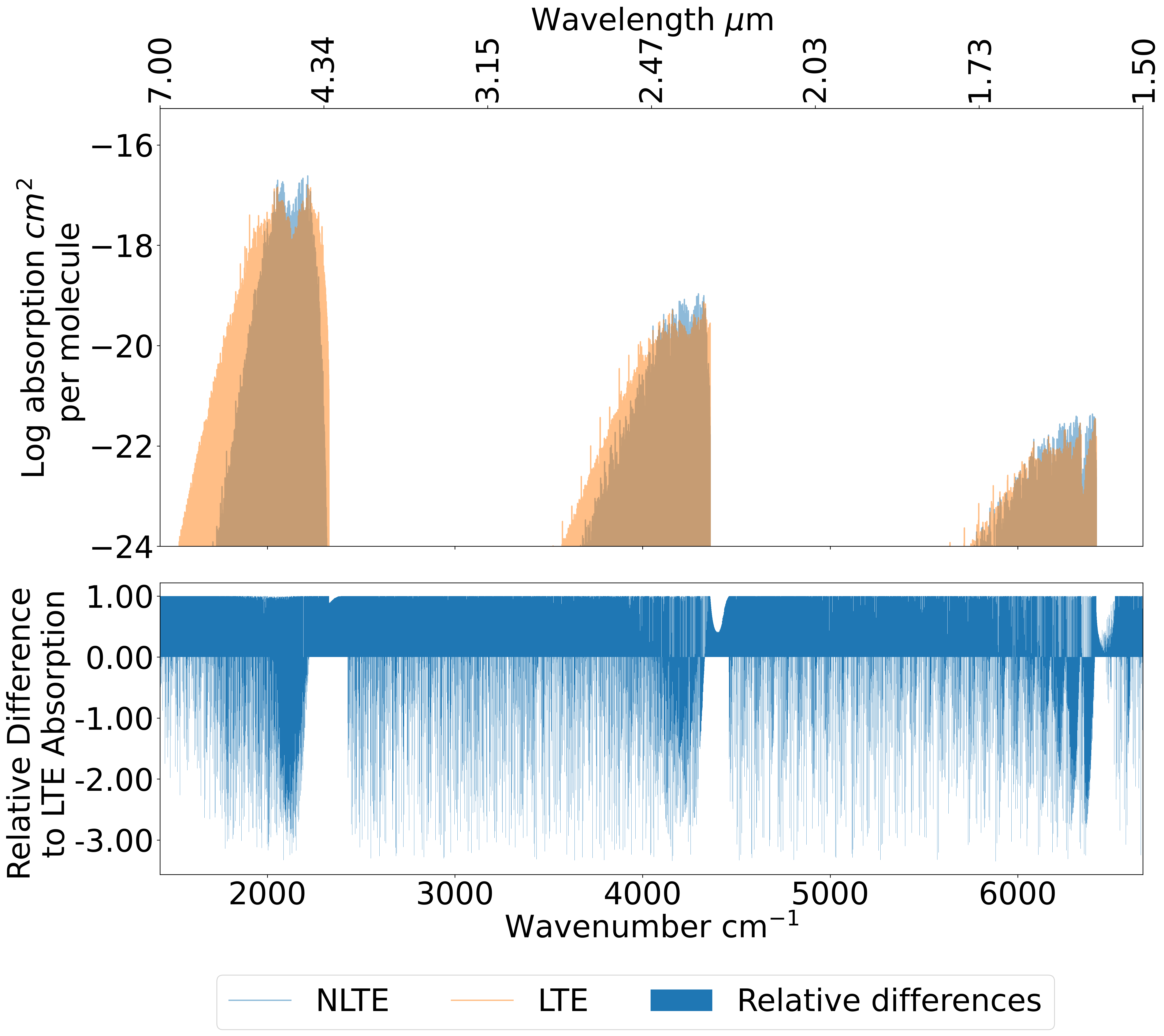}\caption{Opacity cross sections for CO in LTE and non-LTE at WASP-12b's equilibrium temperature (1864~K) and a pressure of 10$^{-4}$ bar at R $\sim$ 150,000, with a rotatational temperature of 700~K for the non-LTE case, as we define in scenario 1. \label{fig:CO_Xsec_diffs_Scenario_1_HRLP} }
\end{figure}

To demonstrate that these differences persist at low pressures, such as those in low pressure atmospheric regions where we expect to find persistent non-LTE effects, the CO LTE and non-LTE cases are plotted in figure \ref{fig:CO_Xsec_diffs_Scenario_1_HRLP} for a pressure of 10$^{-4}$~bar \citep{10SwDeGr.exo}. In addition, this plot shows the cross sections at a resolution of 150,000  (in line with the resolving power of the CRIRES+ instrument) to demonstrate the application to high resolution spectroscopy.

\subsection{Methane Cross Sections}

\begin{figure}
\includegraphics[width=\columnwidth]{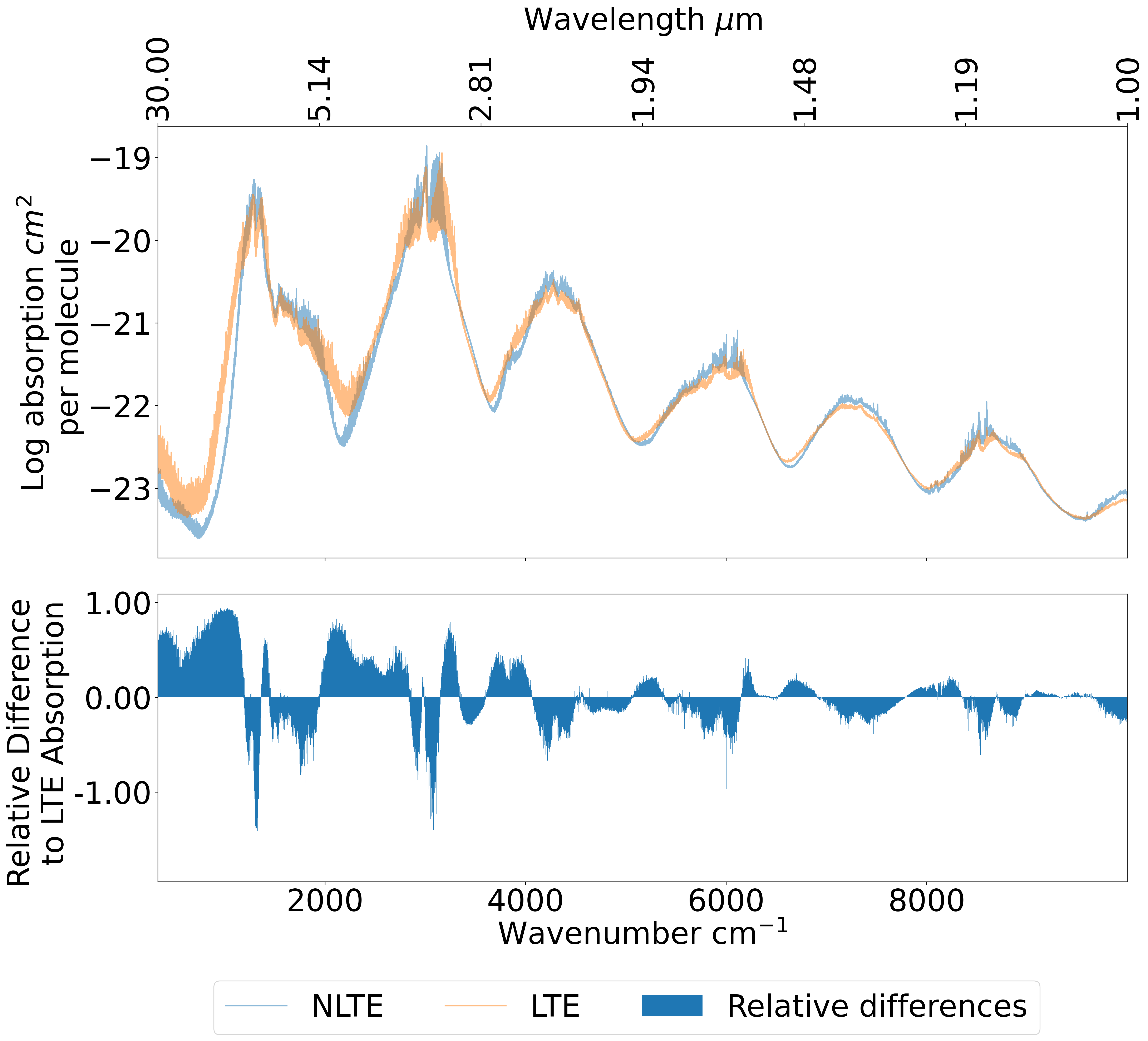}\caption{Opacity cross sections for CH\4\ in LTE and non-LTE at WASP-12b's equilibrium temperature (1864~K) and a pressure of 1 bar, with a rotational temperature of 700~K for the scenario~1 non-LTE case. At R $\sim$ 15000. \label{fig:CH4_Xsec_Rot_Cool_Rel_Diffs} }
\end{figure}

\begin{figure}
\includegraphics[width=\columnwidth]{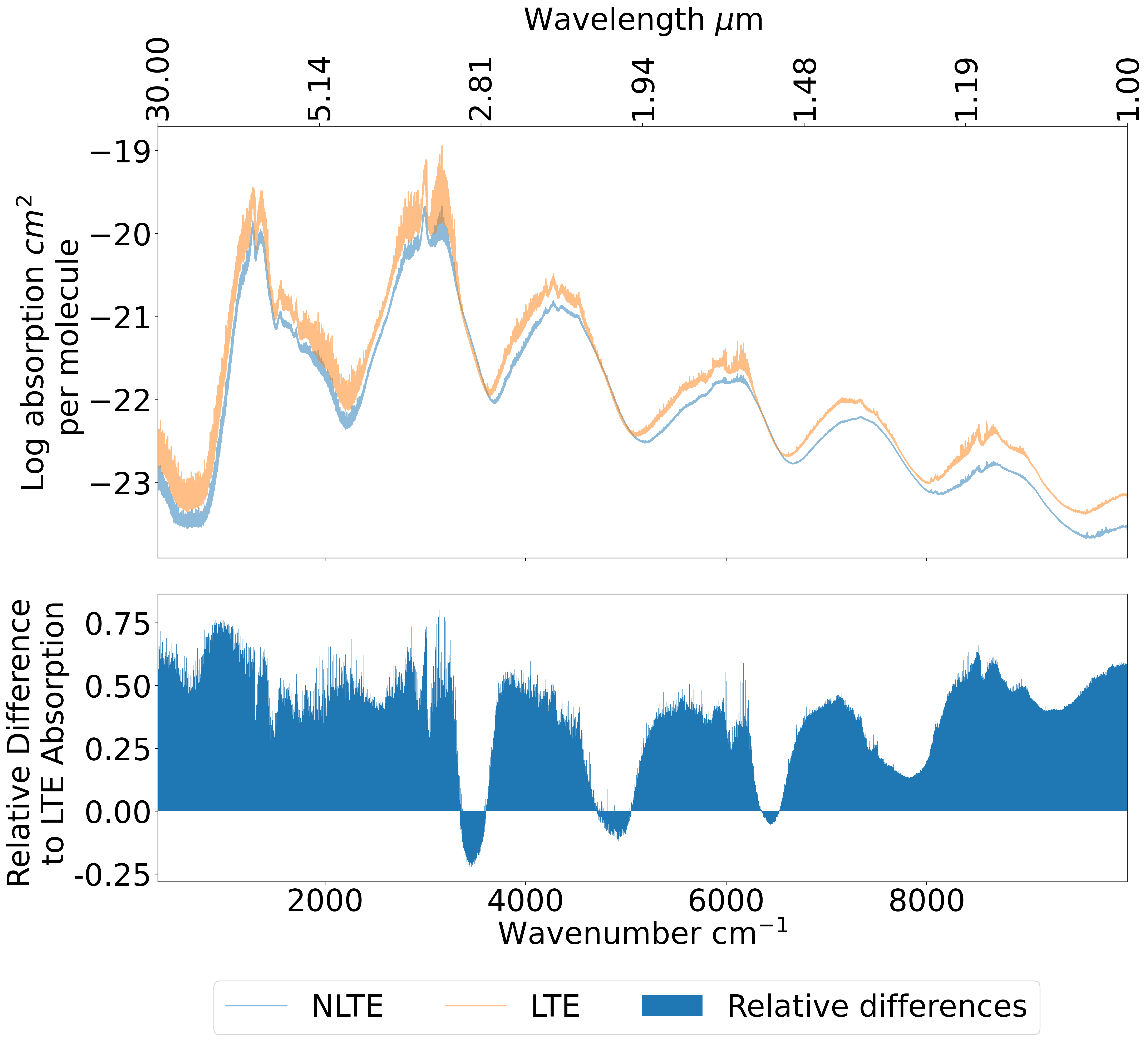}\caption{Opacity cross sections for CH\4\ in LTE and non-LTE at WASP-12b's equilibrium temperature (1864~K) and a pressure of 1 bar, with a vibrational temperature of 3028~K for the scenario 2 non-LTE case. At R $\sim$ 15000. \label{fig:CH4_Xsec_Vib_Excite_Rel_Diffs} }
\end{figure}

The non-LTE opacity cross-sections were generated using the ExoMol CH\4\ line list  10to10 \citep{jt564} at 1 bar of pressure, a resolution of $\sim$ 15,000.
Figure \ref{fig:CH4_Xsec_Rot_Cool_Rel_Diffs} shows the absorptions and differences between them for Methane in LTE ($T= 1864$~K) and non-LTE under non-LTE scenario~1 (700~K/1864~K). In the mid infrared we see relative differences in intensity between 1.8x for the band around 3~\um. When considering wavelengths above 5~$\mu$m, such as will be probed by JWST's MIRI instrument, these relative differences in intensity climb as high as 1.55x and there are substantial wavelength offsets which differentiate the LTE and non-LTE cases.  Substantial differences can also be seen under scenario 2 as in figure \ref{fig:CH4_Xsec_Vib_Excite_Rel_Diffs}, the absorption damping relative to the LTE case gives rise to persistent differences between the two sets of cross sections, most notably above 2.8 \um.

\begin{figure}
\centering
\includegraphics[width=\columnwidth]{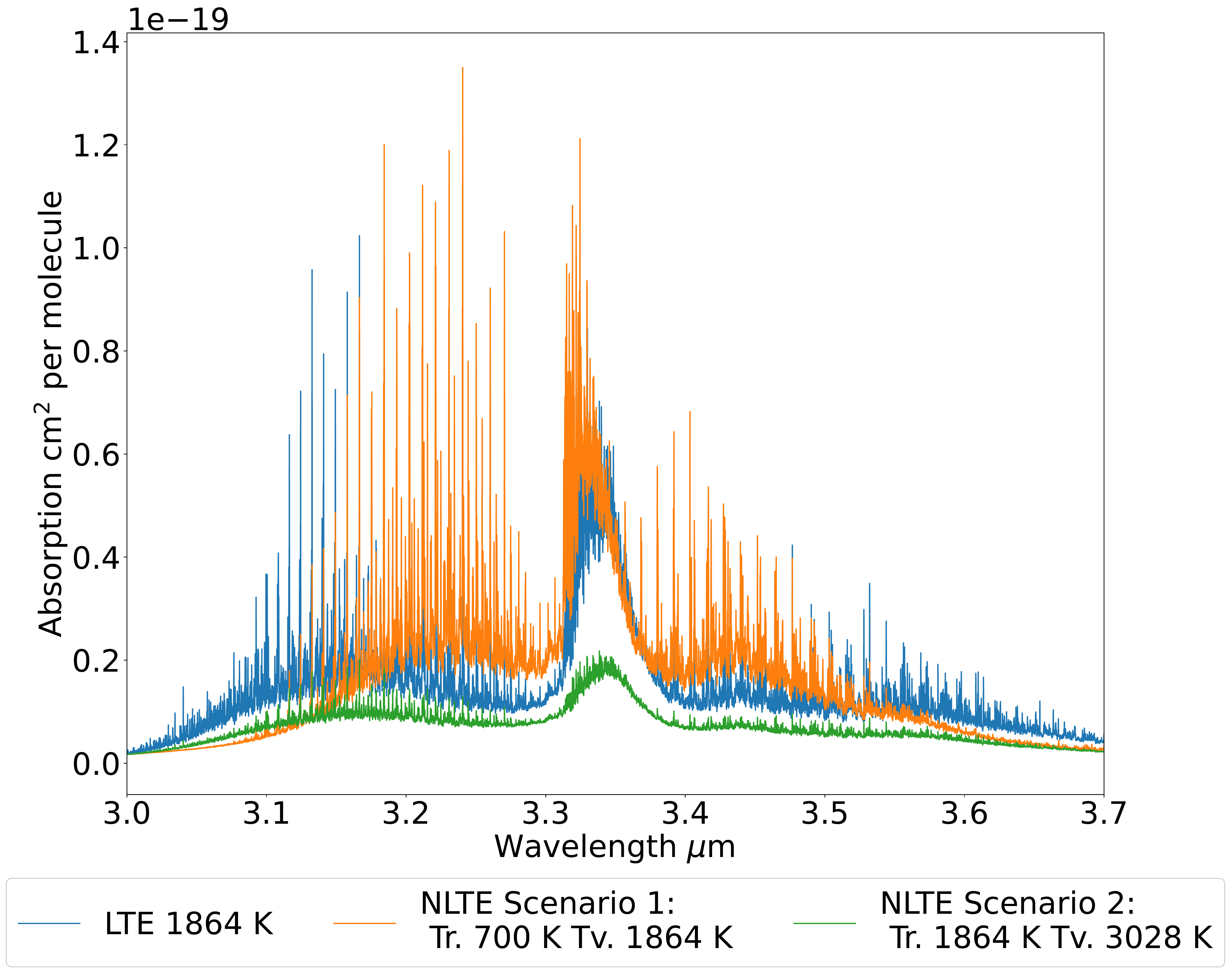}\caption{Opacity cross sections for the band centred around 3.3~\um\ of CH\4\ in LTE and non-LTE at WASP-12b's equilibrium temperature (1864~K) and a pressure of 1 bar, with a rotational temperature of 700~K for the scenario~1 non-LTE case and a vibrational temperature of 3028~K for the scenario 2 non-LTE case. Shown at R $\sim$ 15000. \label{fig:CH4_3mu} }
\end{figure}

Isolating the band centred around $\sim$ 3.3~\um\ in figure \ref{fig:CH4_3mu} shows differences around this band peak; for the scenario~1 non-LTE case this is characterised by a tighter profile when compared to LTE. At peak absorption this can be seen in a wavelength offset of 0.1~\um. The $R$ branch is blue-shifted in the case of non-LTE by $\approx$ 0.0453~\um. Under the second scenario, the non-LTE case is damped with lower absorption than the LTE case.

\begin{figure}\centering
\includegraphics[width=\columnwidth]{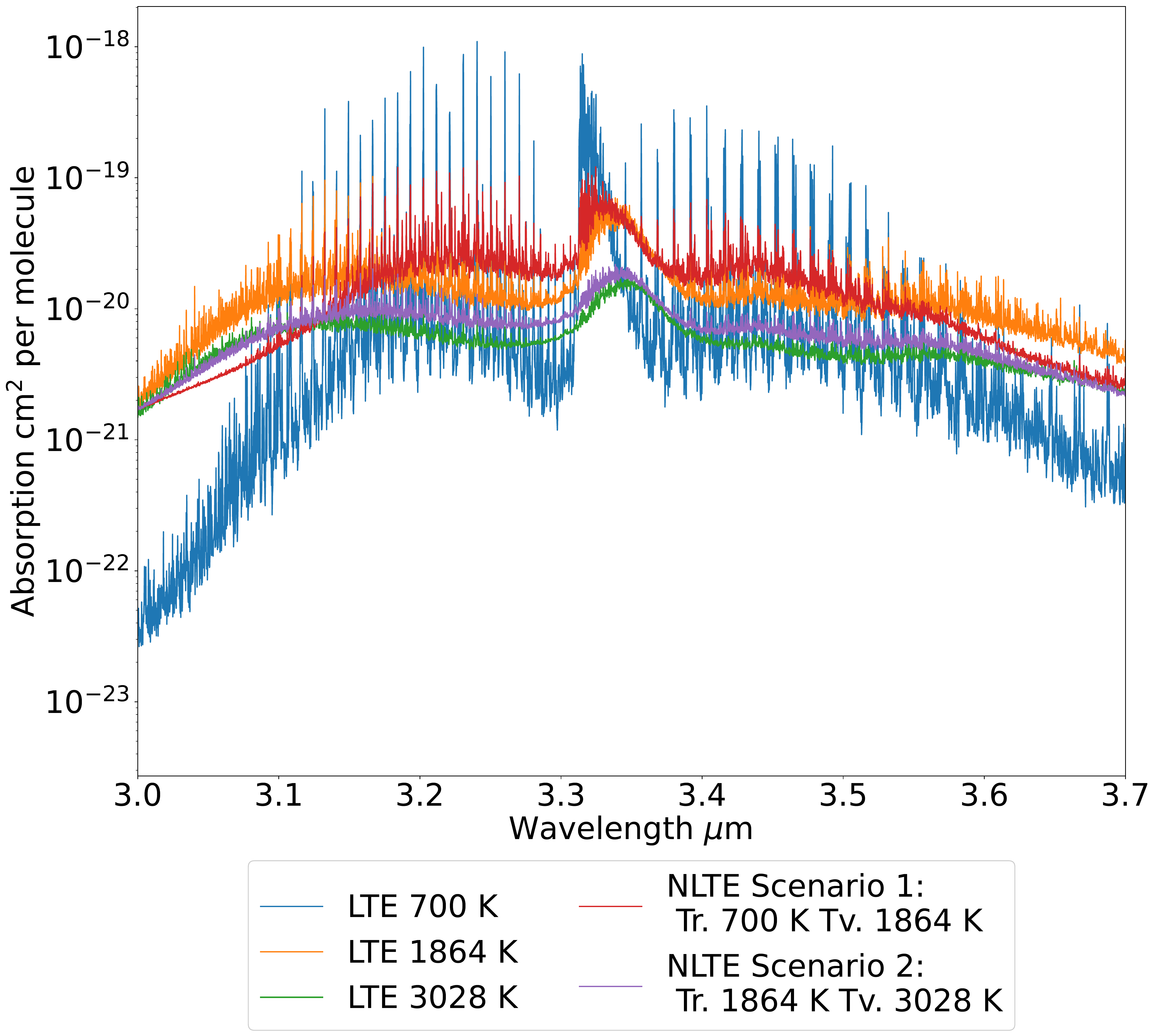}\caption{Opacity cross sections of CH\4\  for the two non-LTE scenarios for the band centered around 3.3~\um\  at R $\sim$ 15000. In addition, three LTE baselines are shown, one for each temperature used in generating the cross sections. \label{fig:CH4_3mu_3_Baselines} }
\end{figure}

It can be seen from figure \ref{fig:CH4_3mu_3_Baselines} that, regardless of the reference temperature chosen (be it the equilibrium temperature or the cooled rotational temperature or the excited vibrational temperature), there are still noticeable differences between both non-LTE scenarios and each LTE baseline.

\begin{figure*}
\includegraphics[width=0.48\textwidth]{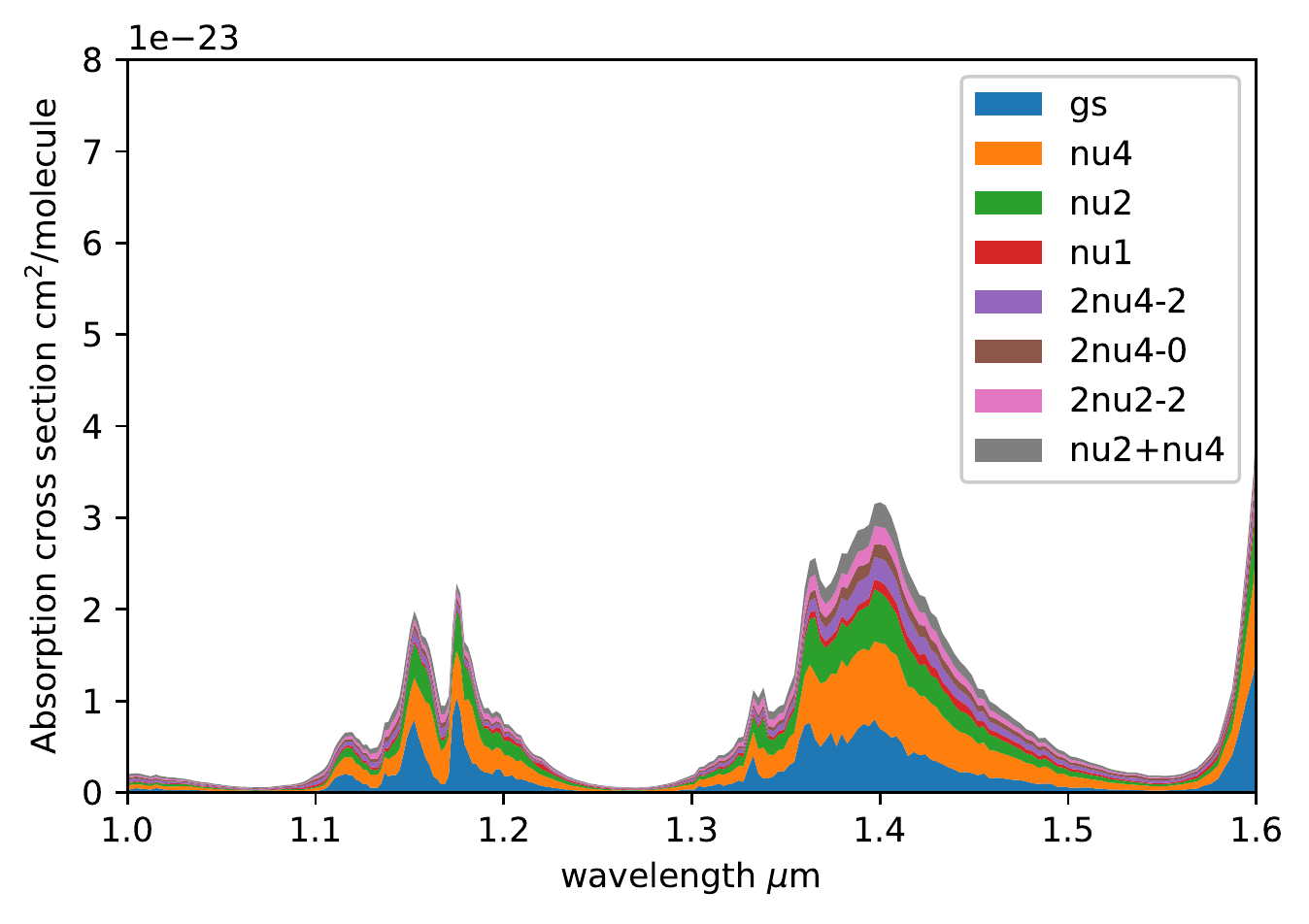}
\includegraphics[width=0.48\textwidth]{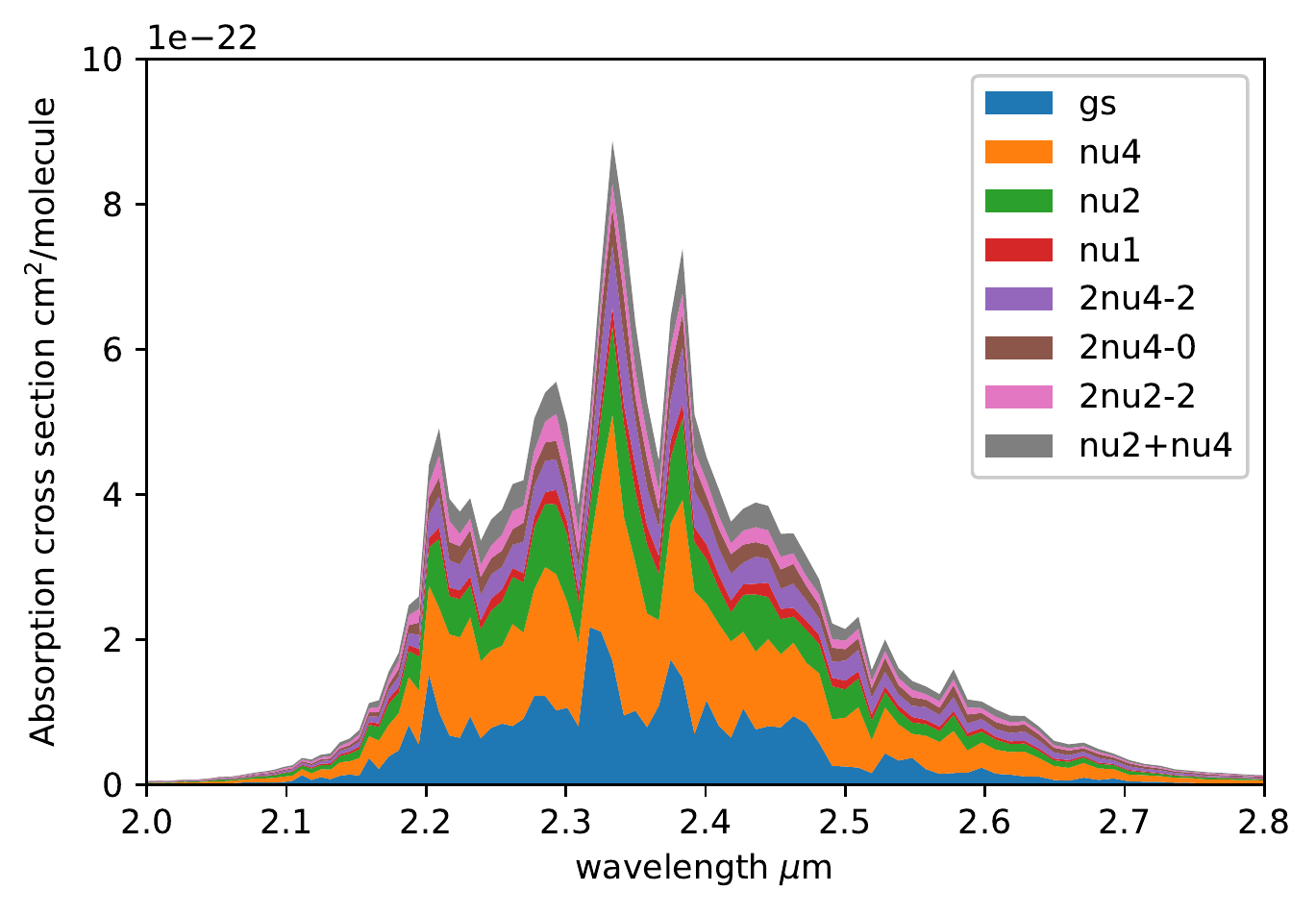}\\ 
\includegraphics[width=0.48\textwidth]{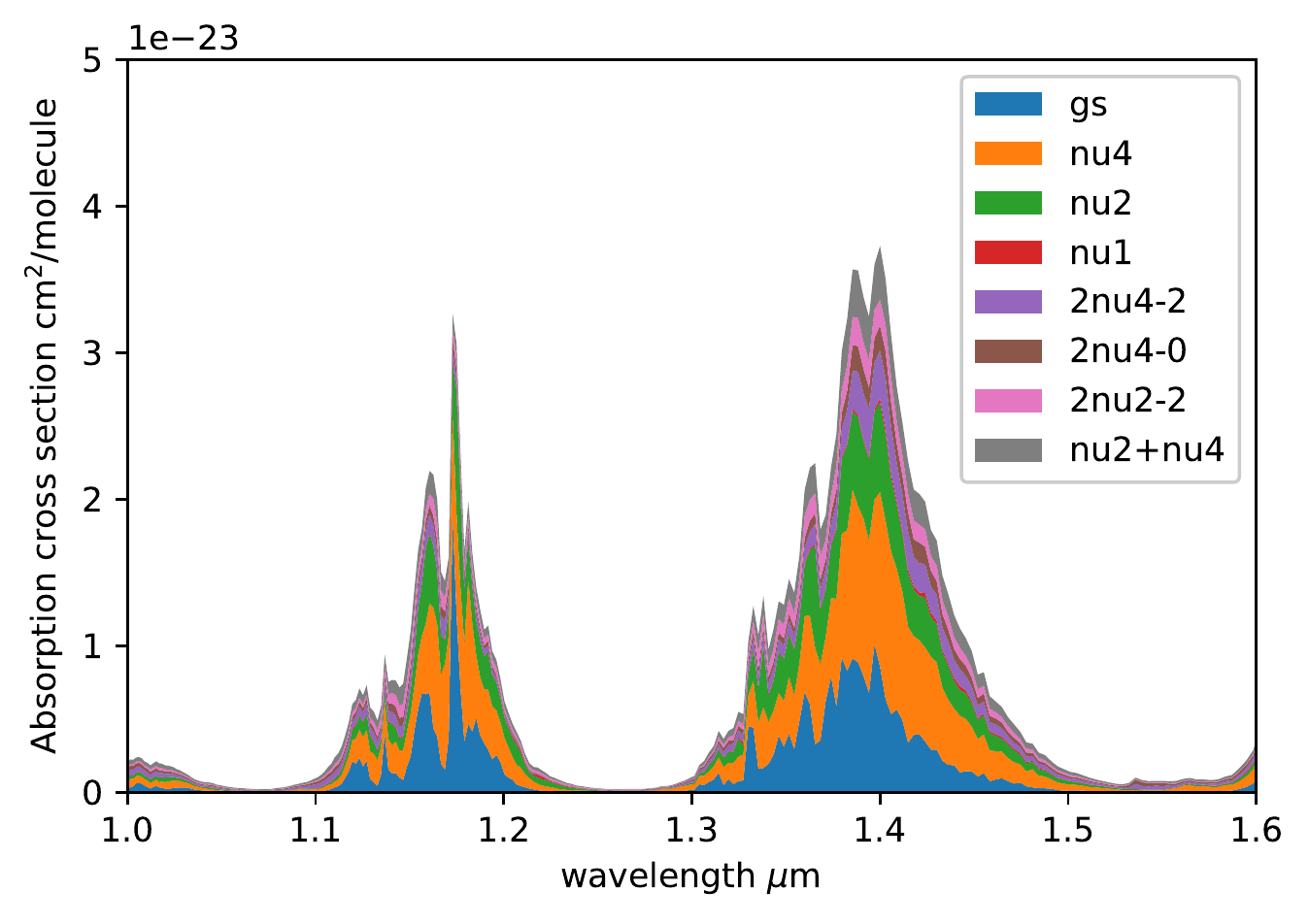}
\includegraphics[width=0.48\textwidth]{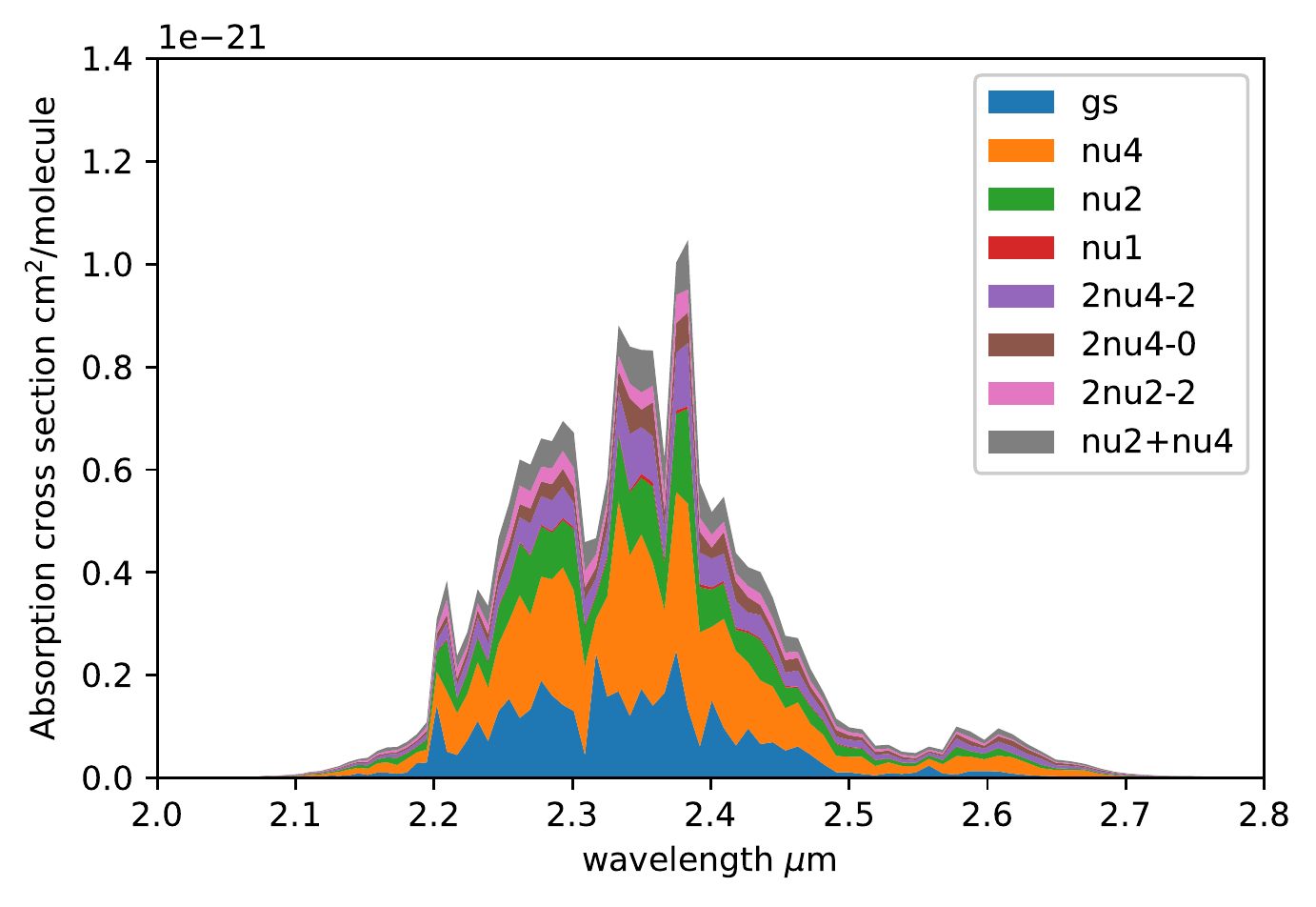}\\
\caption{ \label{fig:CH4_bands}
The individual contributions to the  1.18, 1.4 and 2.4 $\mu$m systems of CH\4\ from different absorption bands, LTE and non-LTE (top two and bottom two panes, respectively). Legends indicate lower vibrational states.
}

\end{figure*}

Figure~\ref{fig:CH4_bands} offers a detailed illustration of a non-LTE spectrum of CH\4\ as formed from individual vibrational bands. The main contribution to the LTE spectrum comes from fundamental/overtone bands (i.e. vibrational bands originating from the ground vibrational state) as well as from the singly excited bending modes $\nu_2$ (doubly degenerate)  and $\nu_4$ (triply degenerate). The latter are characterized by relatively low energies (1533.3 \cm\ and 1310.8~\cm, respectively).  The elevated non-LTE vibrational populations of the other excited states correspond to the elevated vibrational temperature (here $T_{\rm vib} = 1864$~K)  provide a sizeable change to  spectral composition and consequently to the overall spectral shape. The strongest features in non-LTE appear sharper and even shift their positions, see figure~\ref{fig:CH4_bands}. Since the spread of different bands is defined by intensities of the individual rotational lines and  therefore controlled by the same rotational temperature.  When placed on top of each other, as in the case of CH\4\ or H\2O (figure~\ref{fig:H2O_bands}), the hot bands increase the vertical magnitude of the spectral shape while leaving the horizontal spread practically unchanged. This appears to be a general property of the (IR) spectra of non-linear polyatomic molecules, with their hot bands growing on top of each other. 

\begin{figure}\centering
\includegraphics[width=\columnwidth]{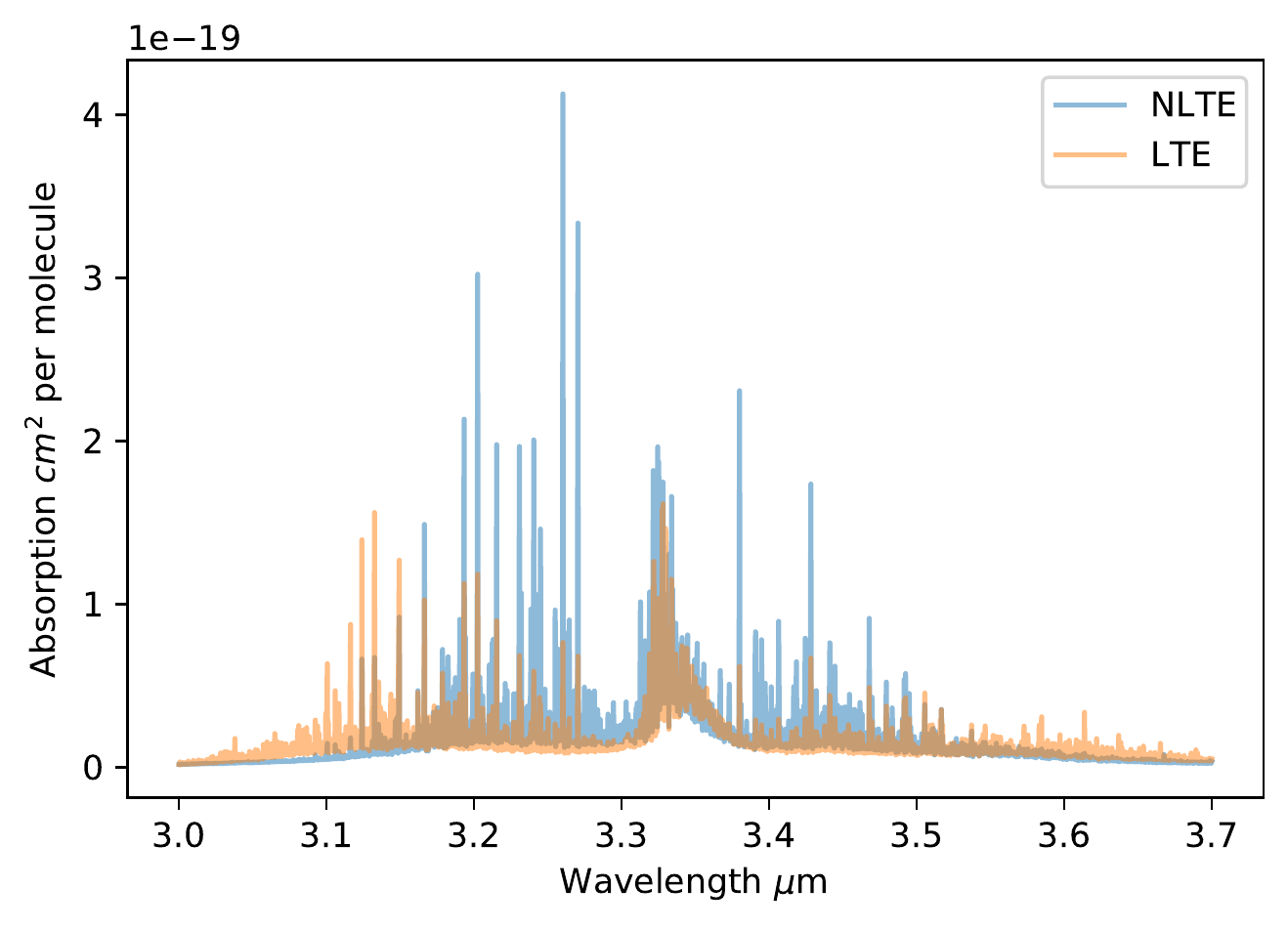}\caption{Opacity cross sections of CH\4\ for the first non-LTE scenarios for the band centered around 3.3~\um~at R $\sim$ 150,000 at a pressure of $10^{-4}$ bar. \label{fig:CH4_3mu_HRLP_Scenario_1} }
\end{figure}

\begin{figure}\centering
\includegraphics[width=\columnwidth]{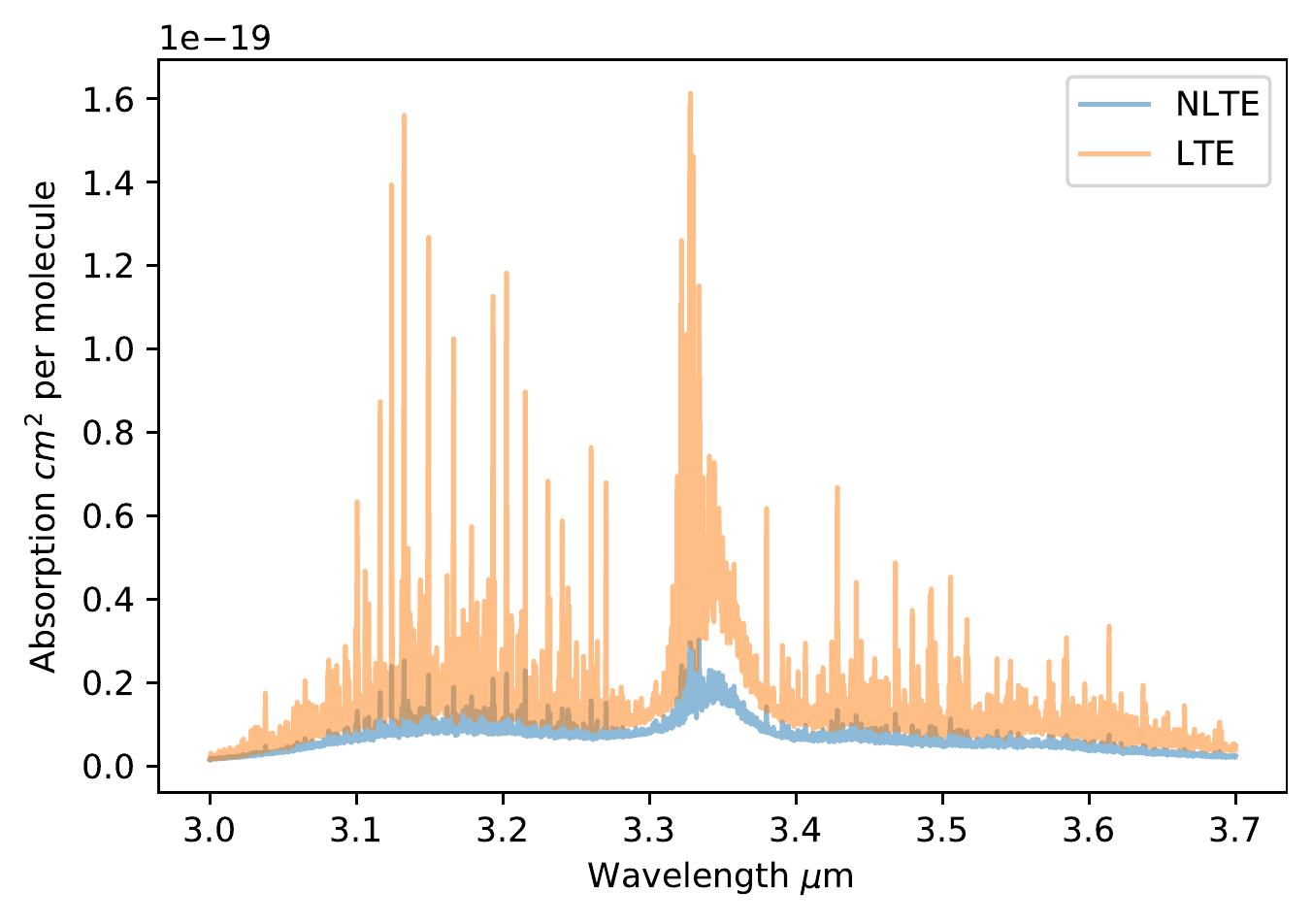}\caption{Opacity cross sections of CH\4\ for the second non-LTE scenarios for the band centered around 3.3~\um~at R $\sim$ 150,000 at a pressure of $10^{-4}$ bar. \label{fig:CH4_3mu_HRLP_Scenario_2} }
\end{figure}

\begin{figure}\centering
\includegraphics[width=\columnwidth]{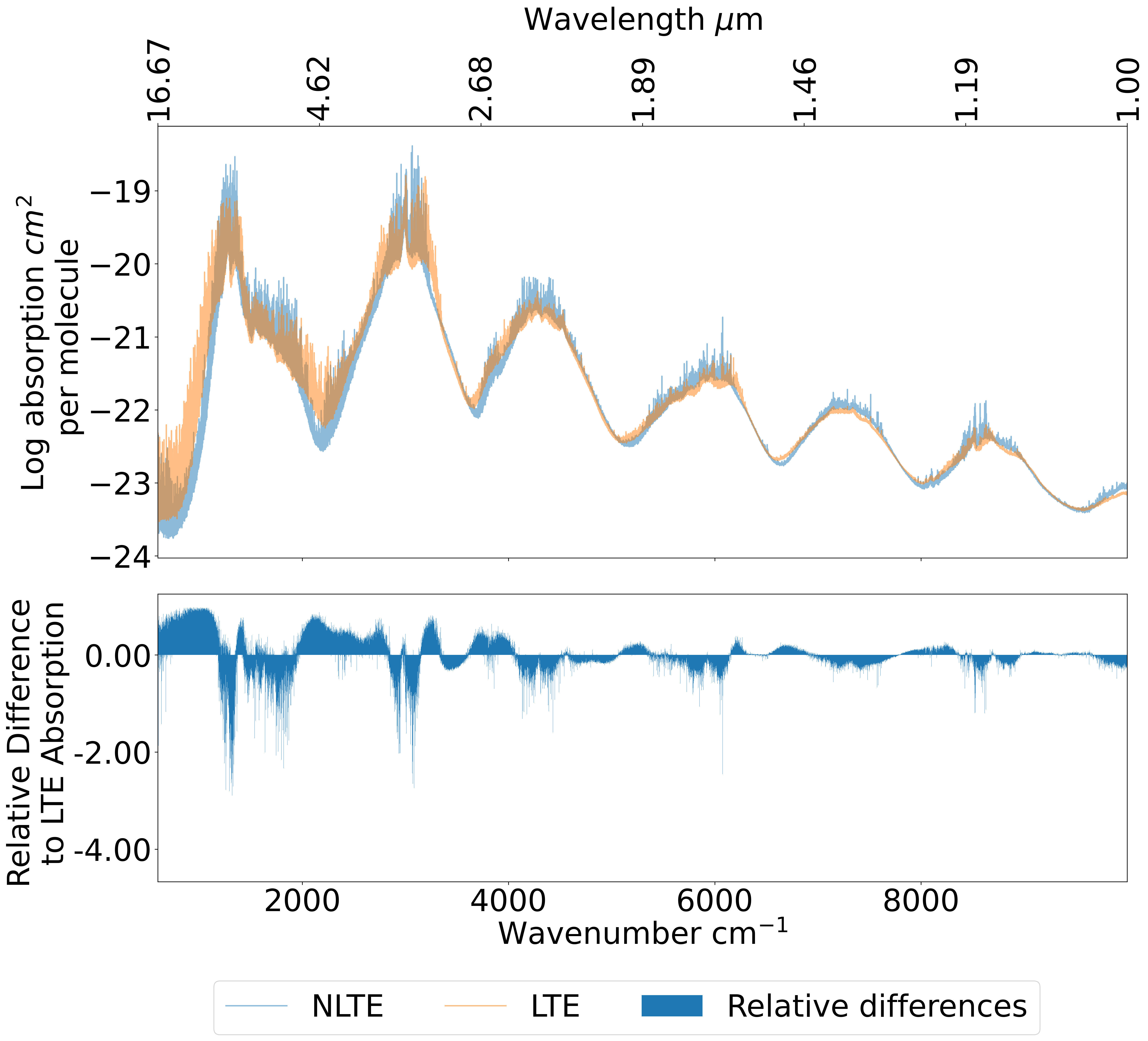}\caption{Differences in opacity cross sections of CH\4\ for the first non-LTE scenarios for the band centered around 3.3~\um~at R $\sim$ 150,000 at a pressure of $10^{-4}$ bar. \label{fig:CH4_3mu_HRLP_Scenario_1_diffs} }
\end{figure}

\begin{figure}\centering
\includegraphics[width=\columnwidth]{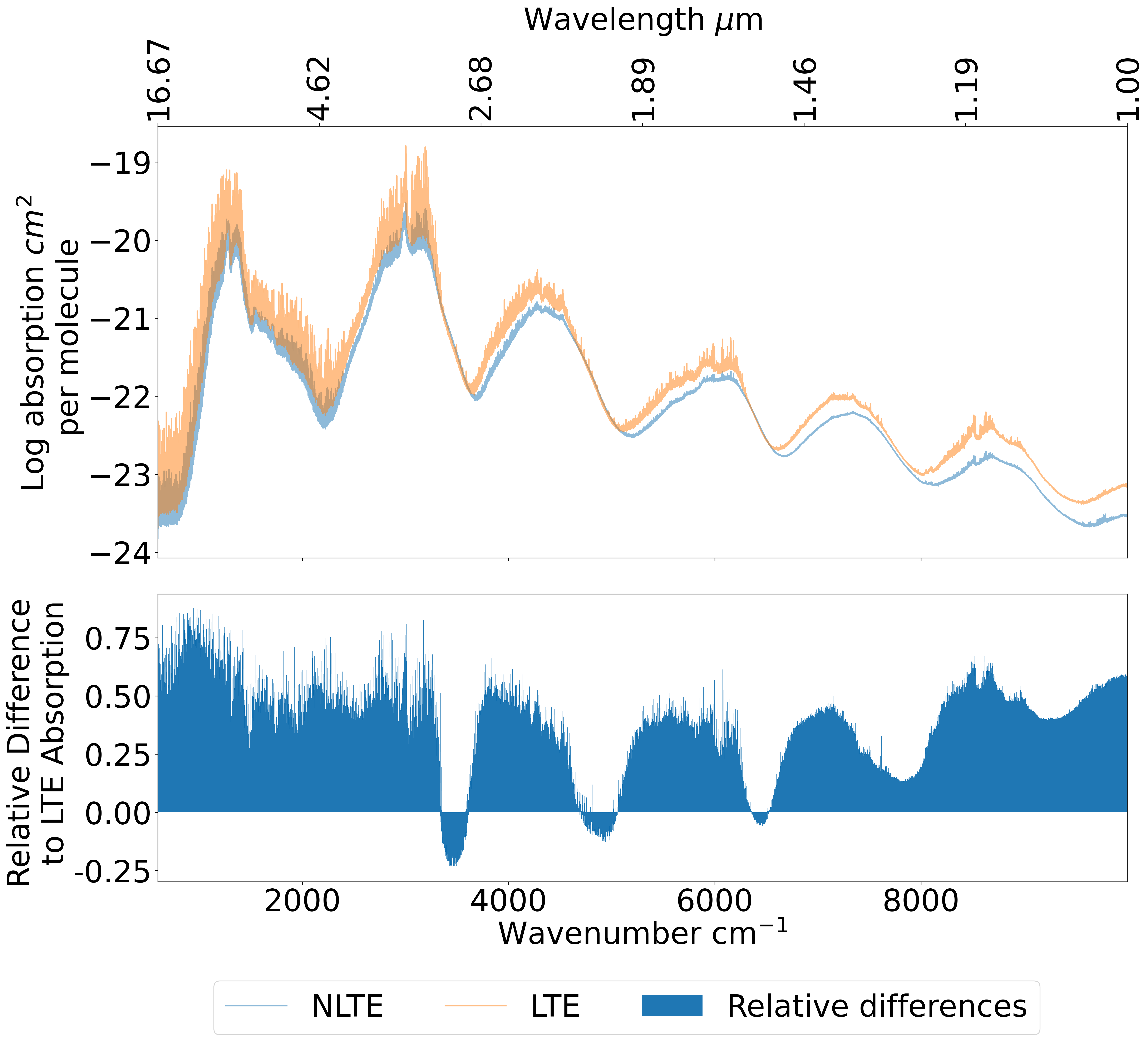}\caption{Differences in opacity cross sections of CH\4\ for the second non-LTE scenarios for the band centered around 3.3~\um~at R $\sim$ 150,000 at a pressure of $10^{-4}$ bar. \label{fig:CH4_3mu_HRLP_Scenario_2_diffs} }
\end{figure}

To demonstrate the presence of these differences at the low pressures where these non-LTE effects occur, the cross sections for CH\4 are plotted for a pressure of $10^{-4}$ bar in figures \ref{fig:CH4_3mu_HRLP_Scenario_1} and \ref{fig:CH4_3mu_HRLP_Scenario_2} for scenario~1 and scenario~2, respectively. These differences are quantified in figures \ref{fig:CH4_3mu_HRLP_Scenario_1_diffs} and \ref{fig:CH4_3mu_HRLP_Scenario_2_diffs}. In addition, these plots show the absorption intensity across wavelength at a resolution of R$\sim$150,000, as may be required for high resolution applications.

\subsection{TiO Cross Sections}

\begin{figure}
\centering
\includegraphics[width=\columnwidth]{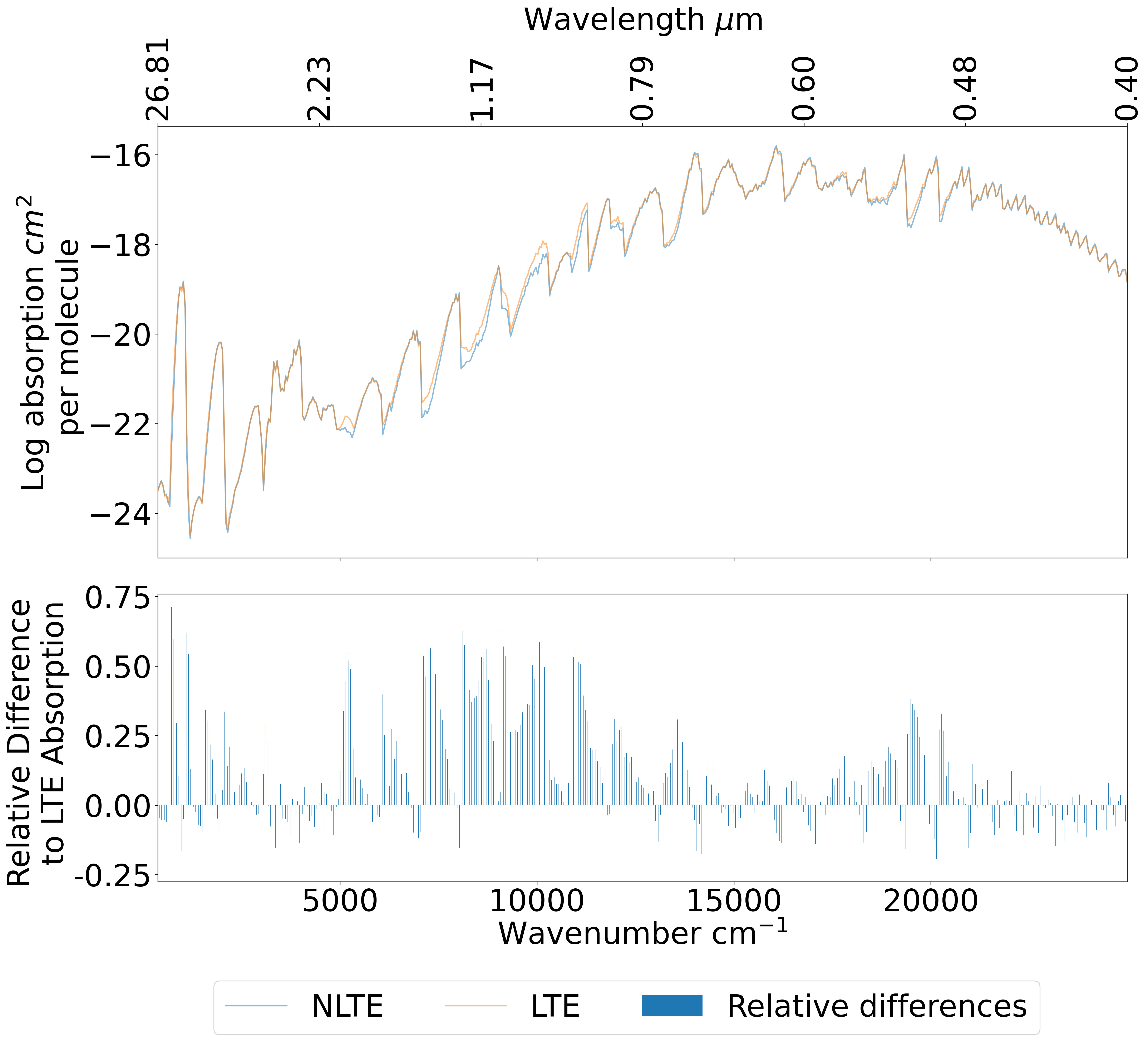}\caption{Absorption cross sections for TiO in LTE and non-LTE at a temperature of 2231~K (with 1800~K for rotational temperature in the non-LTE case) and a pressure of 1 bar and R $\sim$ 150. Following our first scenario, with slight rotational cooling of 431K. \label{fig:TiO_Xsec_Scenario_1_Rel_Diffs}}
\end{figure}

TiO has strong bands in the visible and thus presents a good opportunity to identify the non-LTE effects in the optical, specific for electronic spectra of  (diatomic) molecules. For TiO this can be seen in figure \ref{fig:TiO_Xsec_Scenario_1_Rel_Diffs}, where more persistent and greater differences between the LTE and non-LTE cases exist; reaching up to 1.89x.

\begin{figure}
\centering
\includegraphics[width=\columnwidth]{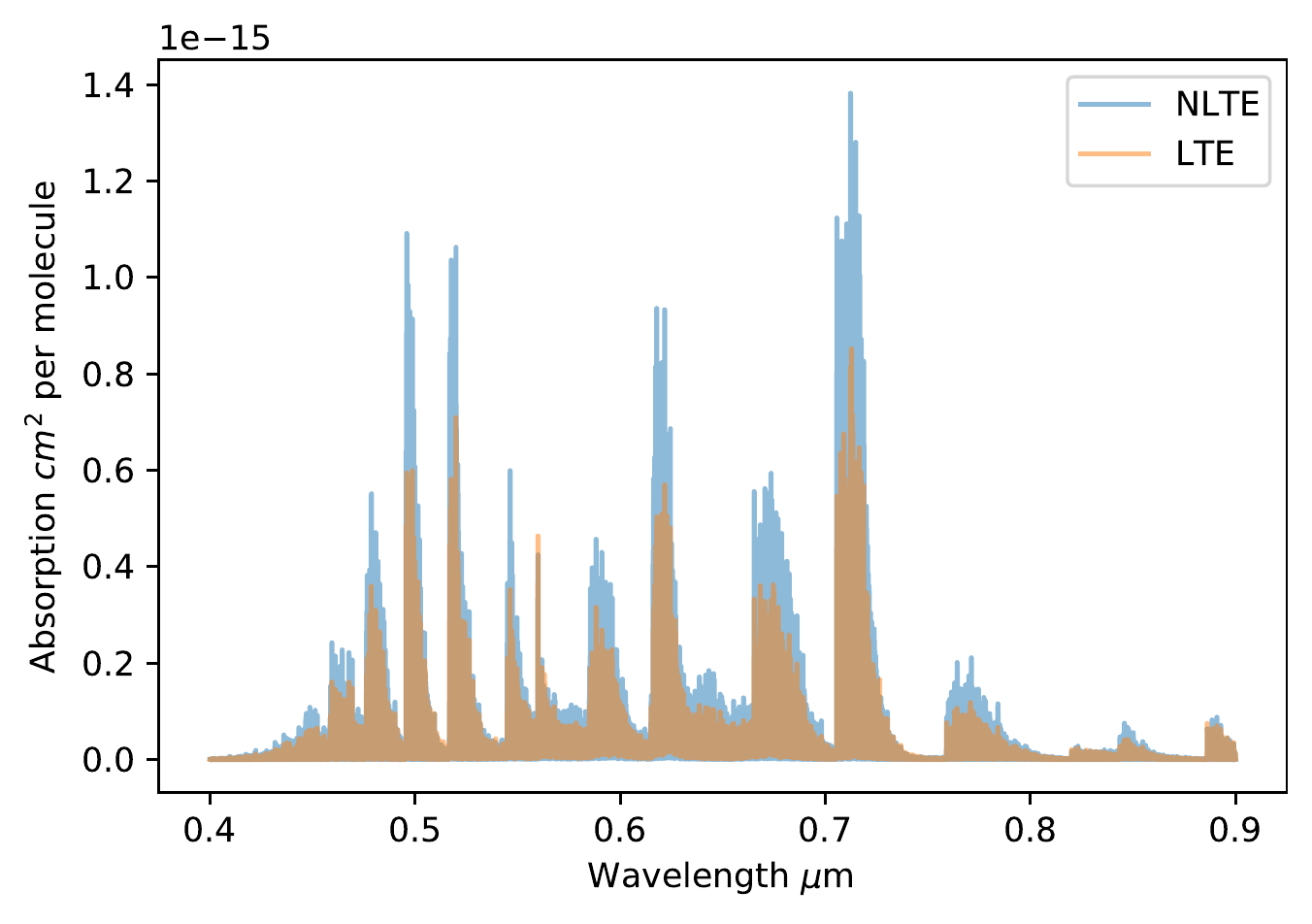}\caption{Absorption cross sections for TiO in LTE and non-LTE at a temperature of 2231~K (with 1800~K for rotational temperature in the non-LTE case) and a pressure of 1 bar with R $\sim$ 150,000 plotted in the visible. Following our first scenario, with slight rotational cooling of 431K. \label{fig:TiO_Vis_Scenario_1}}
\end{figure}

The optical range is shown for non-LTE scenario~1 in figure \ref{fig:TiO_Vis_Scenario_1} where substantial differences in absorption can be seen, with a strong absolute level absorption of the order $10^{-15}$ cm$^{2}$ per molecule. The non-LTE bands here absorb far more strongly in the peaks than the LTE bands and incorporate wavelength offsets leading to wider bases. \ref{fig:TiO_Xsec_Scenario_1_Rel_Diffs}.

\begin{figure}
\centering
\includegraphics[width=\columnwidth]{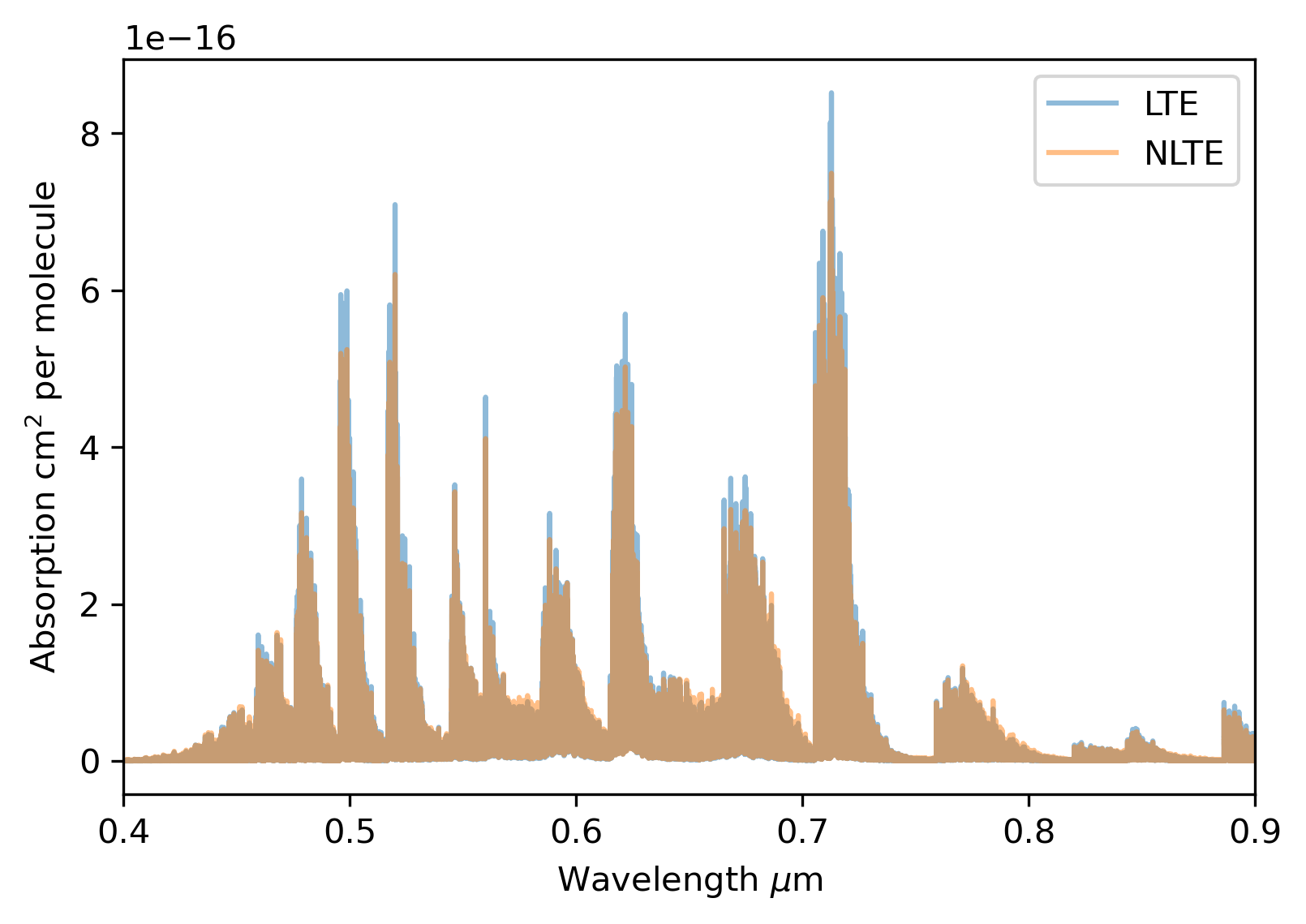}\caption{Absorption cross sections for TiO in LTE and non-LTE at a temperature of 2231~K (with 2662~K for vibrational temperature in the non-LTE case, corresponding to scenario 2) and a pressure of 1 bar and R $\sim$ 150000.  \label{fig:TiO_Xsec_Scenario_2}}
\end{figure}

\begin{figure}
\centering
\includegraphics[width=\columnwidth]{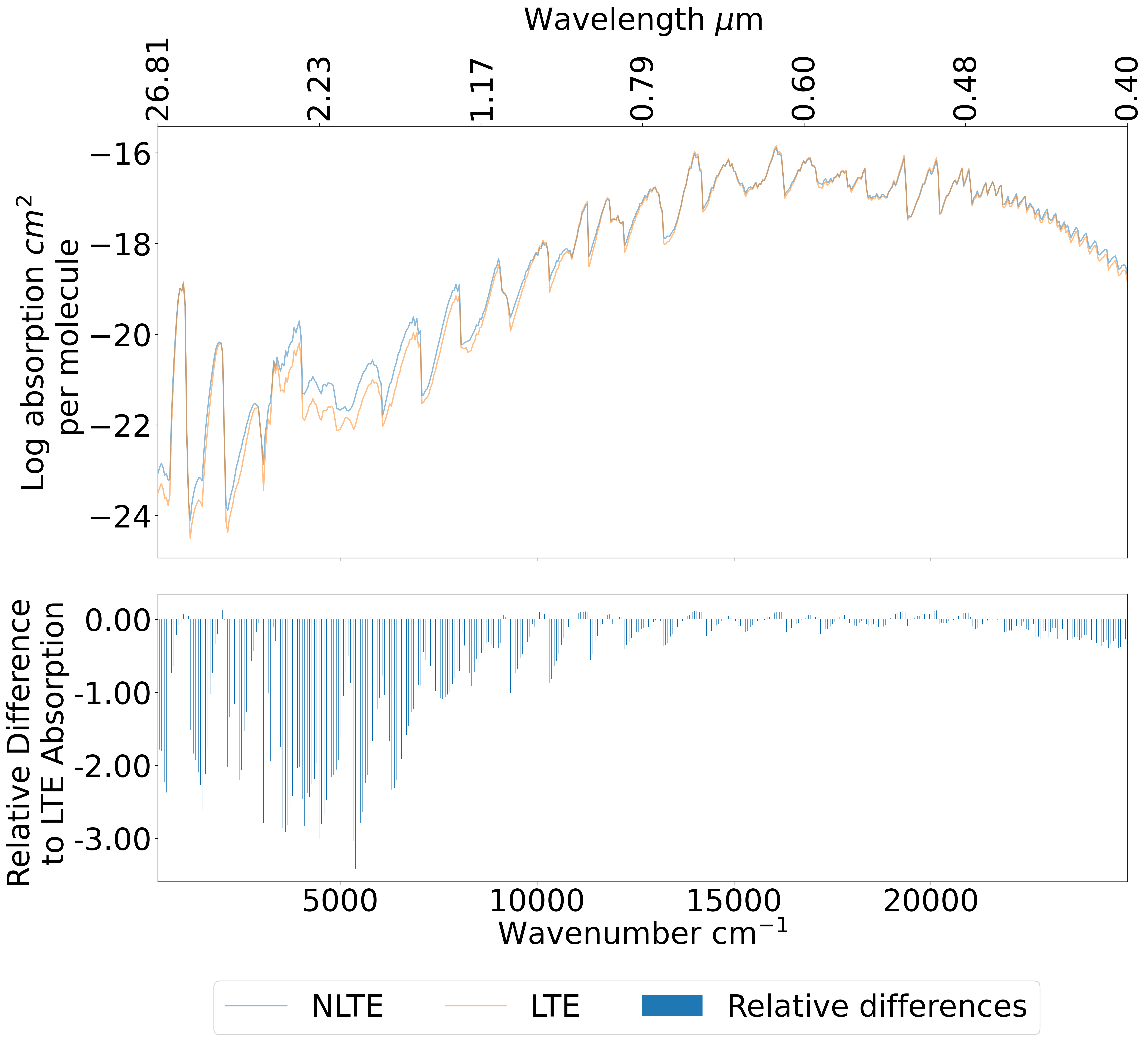}\caption{Absorption cross sections for TiO in LTE and non-LTE at a temperature of 2231~K (with 2662~K for vibrational temperature in the non-LTE case, corresponding to scenario 2) and a pressure of 1 bar with R $\sim$ 150 plotted in the visible. \label{fig:TiO_Vis_Diffs_Scenario_2}}
\end{figure}

\begin{figure}
\centering
\includegraphics[width=\columnwidth]{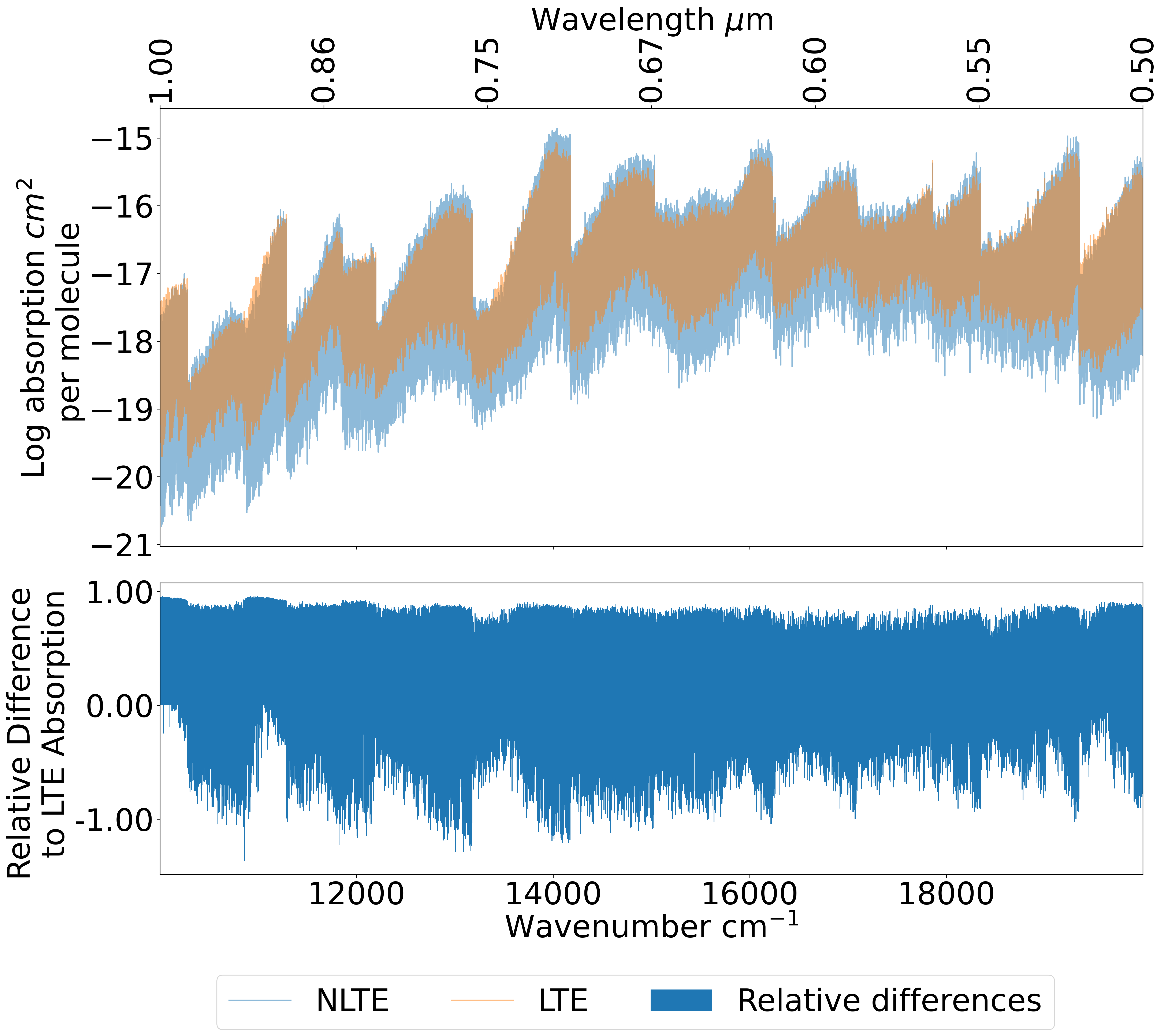}\caption{Absorption cross sections for TiO in LTE and non-LTE at a temperature of 2231~K (with 1800~K for rotational temperature in the non-LTE case, corresponding to scenario 1) and a pressure of 1 bar with R $\sim$ 150,000 plotted between 0.5~\um\ and 1~\um\ . \label{fig:TiO_Vis_Diffs_Scenario_1_R150K}}
\end{figure}

\begin{figure}
\includegraphics[width=\columnwidth]{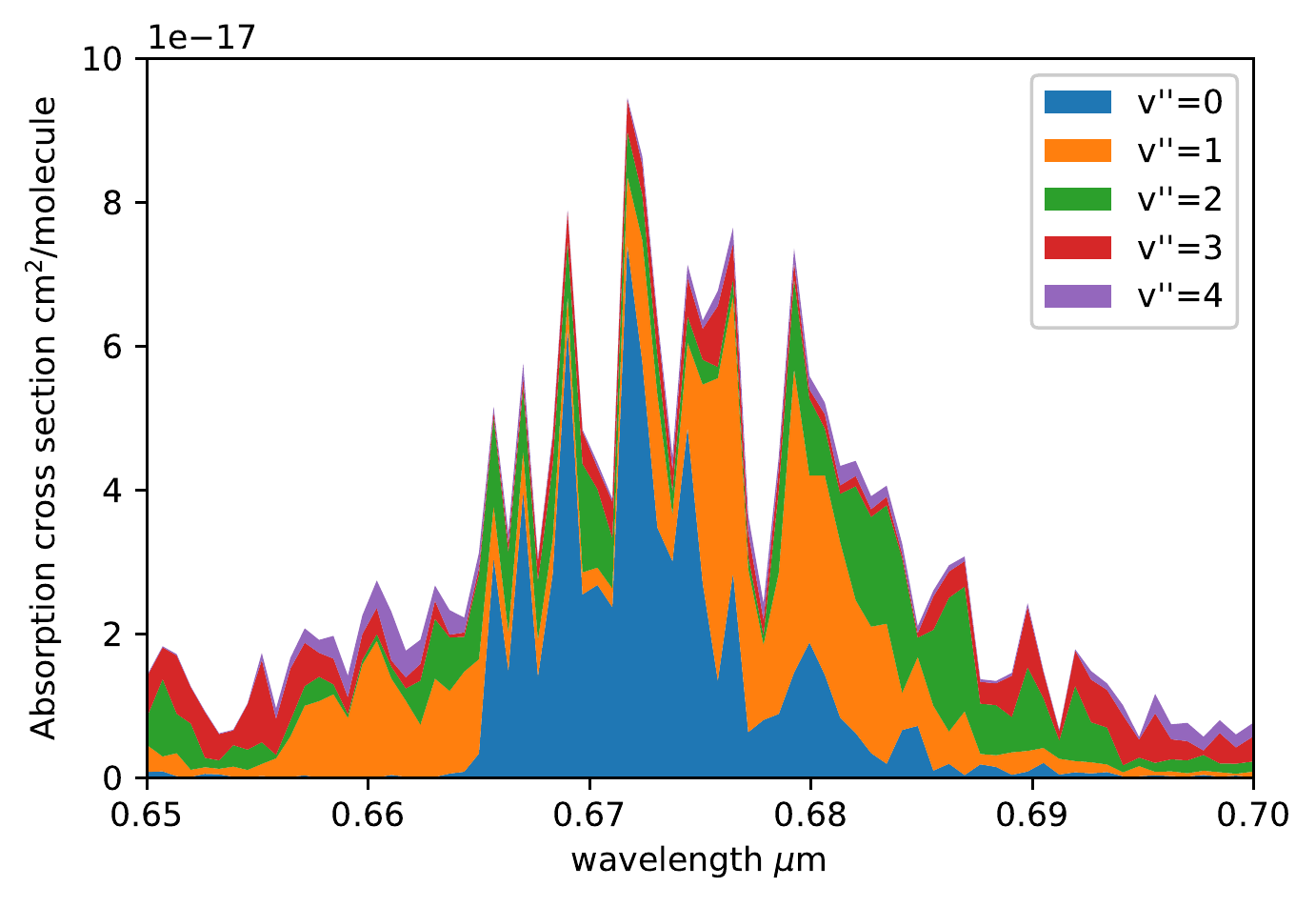} \\
\includegraphics[width=\columnwidth]{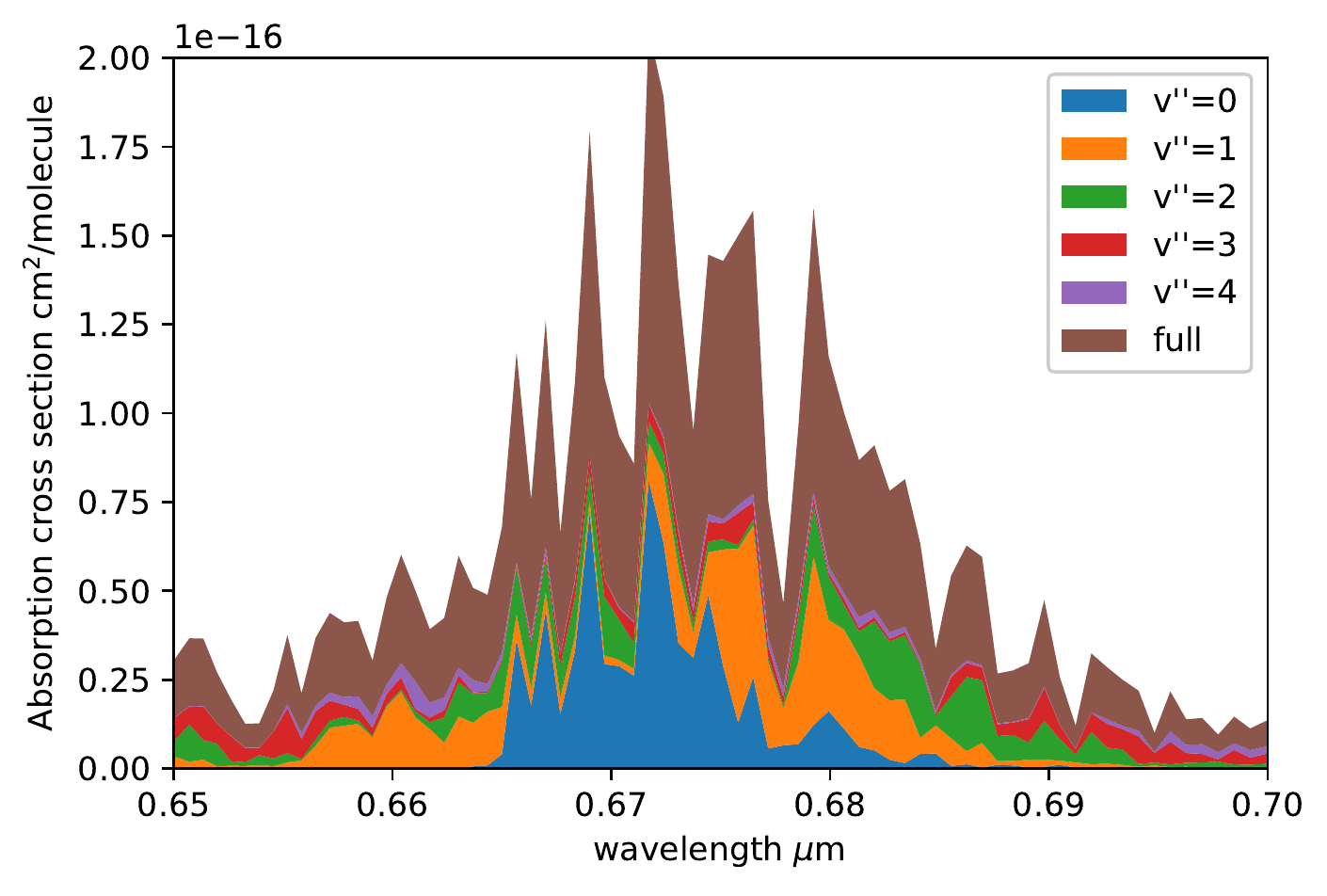} 
\caption{ \label{fig:TiO_bands}
The individual contributions to the $\gamma$ ($A$~$^3\Phi$ $\gets$  $X$~$^3\Delta$) and $\gamma'$ ($B$~$^3\Pi$ $\gets$  $X$~$^3\Delta$)  bands system  of TiO from different vibrational bands, LTE and non-LTE (top and bottom plots respectively). $v''$ represents the vibrational excitation of the lower state in the ground electronic state $X$~$^3\Delta$. }
\end{figure}

The visible region of the TiO absorption is dominated by two electronic bands $\gamma$ ($A$~$^3\Phi$ $\gets$  $X$~$^3\Delta$) and $\gamma'$ ($B$~$^3\Pi$ $\gets$  $X$~$^3\Delta$).
In figure~\ref{fig:TiO_bands} we show the individual contributions from different vibronic bands of TiO in this region.  In the LTE absorption spectrum of TiO at $T=2231$~K  the two strongest $\gamma$ sub-bands are  $\varv'=0 \gets \varv''=0$ and  $\varv'=0\gets \varv''=1$. Here the lower state energy of  $\varv''=1 ,X$ is 999.9~\cm which is small enough to be significantly populated at $T=2231$~K, while the population of the next excited state,  $\varv''=2, X$ ($E$= 1990.8~\cm) is negligible. 

For the non-LTE spectrum of TiO at  $T_{\rm rot}=1800$~K and  $T_{\rm vib}=2231$~K,  shown in Fig.~\ref{fig:TiO_bands}, the hot $\gamma$ sub-bands from the lower excited states $\varv''=2, X$ and $\varv''=3$ ($E$= 2972.6~\cm), $X$ have large enough populations to affect the shape of the spectral band at the longer wavelength, with well defined signatures. To demonstrate this effect at a higher resolution, figure \ref{fig:TiO_Vis_Diffs_Scenario_1_R150K} shows a comparison between TiO in LTE and non-LTE scenario~1 at a resolution of R$\sim$150,000. From this plot, it can be seen that differences between the two cases extend to as much as 2.07x in the intensity of the band peaks. 

We should note the recent non-LTE study of thermodynamic and radiative properties  of TiO by \citet{21BaZhLi.TiO} using high level \textit{ab initio} calculations.

\section{Discussion}

In all four molecules shown here: CH$_{4}$, H$_{2}$O, CO and TiO, there are clear divergences between the absorption cross sections of the LTE and the non-LTE bi-temperature scenarios assumed here. Under scenario~1, these are driven by a reduction in intensity magnitude of the rotational lines as a result of the cooled rotational degrees of freedom used in opacity generation.

In the polyatomic molecules discussed here (CH$_{4}$ and H$_{2}$O) this difference manifests as contraction of  `shoulders'  of the fundamentals bands in cross sections for the non-LTE case comparing to non-LTE, as illustrated for example in figures  \ref{fig:H2O_xsec_scenario_1} and \ref{fig:CH4_Xsec_Rot_Cool_Rel_Diffs}. This provides an observable feature that allows the two cases to be spectroscopically distinguished, and where the rotational cooling is sufficient, is even visible by eye. This potential is maximised by focusing on the bands which present the largest differences, for example two spectral regions shown in figure \ref{fig:H2O_xsec_scenario_1}, and inspecting the relative differences for these regions shown in figure \ref{fig:H2O_xsec_1_5_diffs_scenario_1} illustrates the importance of probing deeper into the infrared to observe these effects. This shows the importance of JWST's broader wavelength coverage over Hubble's WFC3 coverage for non-LTE applications. This improved wavelength coverage, coupled with their higher resolving power, JWST and ground based instruments can provide the necessary platforms for spectroscopic studies to distinguish between the LTE and non-LTE as shown here. Analysing the retrieval of non-LTE under this parameterisation lies beyond the scope of this publication, future work however will focus on examining retrieval results; where it will be necessary to examine how well retrievals can disentangle non-LTE bi-temperature parameters from additional modelled atmospheric effects under self-retrieval of the model, as well as the retrievability of the bi-temperature parameters under varying non-LTE conditions. This will include evaluating T-P profile effects, including bi-temperature pressure profiles which pertain to where and what degree non-LTE effects are introduced under this parameterisation.

For the simple diatomic molecule CO considered in this publication, the differences occur as asymmetric offsets around the bands, as shown in figure \ref{fig:CO_Xsec}. Wavelength offsets such as these present promising opportunities to identify these effects in observed spectra. For CO this must be focused on a viable band however for instance under scenario~1, that which is found in the 3 - 6~\um\ range shown in figure \ref{fig:CO_Xsec}.

The electronic spectra of diatomics represent a special case with the non-LTE vibronic bands shifted relative to each other which helps to recognize the non-LTE effects from the appearance of hot satellite bands. This is in contrast to  IR, vibrational bands of polyatomic molecule, with the non-LTE bands appearing in the same region as  the LTE spectral bands and therefore  hidden by them.

These signatures in H\2O persist under forward modelling as well, with a characteristic non-LTE signature being the absence of `shoulders' around the water band, which are otherwise present in the LTE case.

Since the non-LTE cross section modelling process treats the case where rotational and vibrational temperatures are equal as degenerate with the LTE case, it is intuitive that the differences between the two circumstances decrease as the two temperatures converge. A physical effect sufficient to drive a strong enough difference in the energies across these two molecular modes is therefore required to produce a resolvable non-LTE signature in a spectrum observed with the addition of instrumentation error, along with low enough collision rates such that collisions between the molecules do not lead to a strong effective return to local thermodynamic equilibrium. 

A number of physical effects could present significant driving forces of these non-LTE conditions at high altitudes in the atmospheres of hot Jupiters, for instance: stellar pumping from a nearby star or shock regions at the boundaries of fast jet streams. In addition, it is known from laboratory experiments that non-LTE effects are important in bow shocks such as those at the boundaries of planetary atmospheres \citep{20ScBoHa.nLTE}. Large temperature differentials, such as those common between the day and night sides of tidally locked hot Jupiters, can cause very fast jet streams high in the atmosphere \citep{11ShPoxx}. Shock regions form where these jet streams break, with the rotational mode of the molecules cooling preferentially, giving rise to non-LTE conditions; this rotational cooling has been observed experimentally under laboratory conditions. Non-LTE conditions like this are due to the equilibrium between the energies of a molecule's modes breaking down at the normal shock around a supersonic jet \citep{01MaGrEl}. There are also vertical wind shears in the form of supersonic jets, as suggested to be present in the atmosphere of HD\,189733\,b \citep{16BrdeSn,20SeEhPi}; these fast moving winds with shocks occurring at high altitudes with low pressures present a prime opportunity for the formation and persistence of non-LTE conditions in this manner. In the case of strong stellar irradiation, incoming photons from the host star excite the vibrational molecular mode preferentially; this is referred to as `pumping' and may drive molecules to non-LTE states even at lower altitudes \citep{12FeKuxx}. In the laboratory setting, plasma glow discharges have produced vibrational excitation in CO\2\ \citep{17KlEnBe.nLTE} where the vibrational temperature is recorded at 550\,K over the rotational temperature.
To name a few examples within our own solar system, in Earth's atmosphere, non-LTE effects are present with visible effects in the form of air-glow emissions, as noted for vibrationally excited OH in \citep{71ChMaPo.OH}, for which level populations have been measured \citep{20NoWiGo.OH}. In the atmosphere of Mars, CO\2 has a non-LTE emission feature at 4.3~\um\ \citep{05LoLoLo} \citep{18GiFoLo.nLTE}; In the atmosphere of Venus non-LTE emissions have been detected in limb observations for CO \mbox{\citep{15GiLoPe.nLTE}} and CO\2 \mbox{\citep{09GiLoDr.CO2}}. For CH\4 \mbox{\citep{11GaLoFu.CH4}} and HCN \mbox{\citep{13ReKuFa.HCN}} in Titan's atmosphere and for CH\4 in the atmospheres of Jupiter and Saturn \mbox{\citep{99DrFoCr.CH4}}.

\section{Conclusion}

A simplified bi-temperature  model for non-LTE spectra of exoplanetary atmospheres is used to investigate features of vibrational bands to produce non-LTE spectral signatures for  four key atmospheric molecules, CH$_{4}$, H$_{2}$O, CO and TiO. These molecules sample different spectra ranges (IR and Vis),  different spectral types (electronic, ro-vibrational) as well as different molecular tyoes (diatomics and polyatomics). In the case of the visible spectrum of the diatomic molecule TiO,   it can be seen that the hot bands are shifted from the main, most-populated band, which makes it easier to distinguish and also to use as a non-LTE indicator.  For the (non-linear) polyatomics, the hot bands sit on top of each, making their recognition more challenging. Still, the shapes of the non-LTE bands show  differences which should be significant enough to make them detectable. 

It can be seen from this investigation that the presence of CH$_{4}$, H$_{2}$O, CO and TiO molecules in a state of non-local thermodynamic equilibrium can be distinguished from their presence in LTE by way of variations in their absorption relative to each other. This underlying variation in the spectroscopic characteristics of the molecules based on the state of their local thermodynamic equilibrium is a method by which such effects could potentially be detected. 

Among this sample of molecules, differences in both the optical and infrared regions of the spectrum can be seen, indicating the potential for detection of these effects with multiple observing methods including JWST and ground-based high resolution facilities.

Across the molecules, a number of prominent signatures emerge to distinguish non-LTE. For instance CH\4\ has a prominent signature where the LTE and scenario~1 non-LTE absorption cross sections diverge around 3.15 ~\um\ - this shift of the $R$ branch gives a distinctive spectral feature by which to distinguish CH\4\ in non-LTE. Under scenario 2, absorption variations exist across the board giving a noticeable difference. While for H\2O, such signatures are exhibited as an absence of shoulders around the 2.0~\um\ and 2.7~\um\ bands for scenario~1. The sole strong Carbon Monoxide signature for scenario~1 non-LTE is visible as an offset between 5 and 6 \um, while additional strong offsets exist for scenario 2 vibrational excitation around the 2.76~\um\ and 1.75~\um\ bands. In the case of TiO numerous strong signatures exist in the optical part of the spectrum, present in the bands distributed between 0.5 and 0.75 \um, most notably for rotational cooling. The nature of the non-LTE manifestation for the electronic spectrum of TiO  is different from the IR spectra of H\2O, CH\4\ or CO: it manifests as intensity variations of the vibronic bands rather than contraction of the band shoulders for the rovibrational spectra. The result is that the non-LTE vs. LTE differences are intuitively visible as a narrowing of the bands for H\2O, CH\4\ and CO, and as higher peaks for TiO.

Non-LTE can, and does, arise frequently in planetary atmospheres in a variety of scenarios. Rather than examine each of these cases individually (if indeed it is tractable), the model presented here describes the aggregate non-LTE effect via the bi-temperature parameterisation which is used in many cases. Taking separate rotational and vibrational temperatures as parameters for a two Boltzmann temperature model provides a robust and efficient approach to introducing this additional degree of freedom with which to model non-LTE spectral effects, while the form of the additional temperature dependence term can be further explored. In this way non-LTE absorption characteristics can be determined for molecules with this two-temperature model, without the need to explicitly calculate non-LTE population densities that require a priori knowledge of specific physical mechanisms within atmospheric scenarios so detailed as to be incompatible with the exoplanet use case.

\section{Further Work}
Further work on this topic will investigate non-LTE effects in high resolution exoplanet spectra. In addition, this non-LTE approach will be investigated as an addition to traditional retrieval frameworks, using the same non-LTE cross section approximations to introduce additional fitted temperature variables. This will seek to quantify the degree to which an observed spectrum can be explained by a given molecule existing in a state of non-LTE. Additional work will look into modelling of the day-night side temperature dynamics to place additional constraint on the differences in molecular rotational and vibrational temperatures caused by physical conditions as well as atmospheric pressure constraints driving molecular populations back to local thermodynamic equilibrium.

\section*{Acknowledgements}
{\it Funding:} SW was supported by the STFC UCL Centre for Doctoral Training in Data Intensive Science (grant number ST/P006736/1)
This project has received funding from the European Research Council (ERC) under the European Union's Horizon 2020 research and innovation programme (grant agreement No 758892, ExoAI) and the European Union's Horizon 2020 COMPET programme (grant agreement No 776403, ExoplANETS A) and the ERC Advanced Grant Project No. 883830. Furthermore, we acknowledge funding by the UK Space Agency and Science and Technology Funding Council (STFC) grants:  ST/R000476/1, ST/K502406/1, ST/P000282/1, ST/P002153/1, ST/S002634/1 and ST/T001836/1. The authors acknowledge the use of the UCL Legion High Performance Computing Facility (Legion@UCL) and associated support services in the completion of this work, along with the Cambridge Service for Data Driven Discovery (CSD3), part of which is operated by the University of Cambridge Research Computing on behalf of the STFC DiRAC HPC Facility (www.dirac.ac.uk). The DiRAC component of CSD3 was funded by BEIS capital funding via STFC capital grants ST/P002307/1 and ST/R002452/1 and STFC operations grant ST/R00689X/1. DiRAC is part of the National e-Infrastructure.

{\it Software:} ExoCross \citep{ExoCross}, TauREx3 \citep{TauREx3}, Numpy \citep{NUMPY}, Pandas, 
h5py, 
Matplotlib \citep{MATPLOTLIB}.

\section*{Data Availability}

 The underlying molecular linelist data from the ExoMol project is available from the Exomol website (\url{www.exomol.com}). Derived non-LTE data is available from the corresponding author upon reasonable request.



\bibliographystyle{mnras}
\bibliography{journals_astro,methods,exoplanets,stars,non-LTE,ISM,Books,NO,programs,jtj,CO,TiO,comets,atmos,AlO,planets,OH,CH4, CO2,HCN} 





\bsp	
\label{lastpage}
\end{document}